\documentclass[%
 reprint,
 amsmath,amssymb,
 aps,
]{revtex4-1}

\pdfoutput=1

\usepackage[english]{babel}
\usepackage{graphicx}
\usepackage{epsfig}
\usepackage{color}
\usepackage{amssymb}
\usepackage{amsmath}
\usepackage{array}
\usepackage[loose]{units}
\usepackage{braket}
\usepackage[caption=false]{subfig}

\newcommand{\UO}{Department of Physics, University of Oregon, Eugene, Oregon}
\newcommand{\Glasgow}{School of Engineering, University of Glasgow, Glasgow G12 8QQ, Scotland, UK}

\begin{document}

\title{Efficient Sorting of Free Electron Orbital Angular Momentum}
\author{Benjamin J. McMorran}õ
\email{mcmorran@uoregon.edu}
\affiliation{\UO}
\author{Tyler R. Harvey}
\affiliation{\UO}
\author{Martin P. J. Lavery}
\affiliation{\Glasgow}

\date{September 24, 2015}
                            
\begin{abstract}
  We propose a method for s6rting electrons by orbital angular momentum (OAM). Several methods now exist to prepare electron wavefunctions in OAM states, but no technique has been developed for efficient, parallel measurement of pure and mixed electron OAM states. The proposed technique draws inspiration from the recent demonstration of the sorting of OAM through modal transformation. We show that the same transformation can be performed with electrostatic electron optical elements. Specifically, we show that a charged needle and an array of electrodes perform the transformation and phase correction necessary to sort orbital angular momentum states. This device may enable the analysis of the spatial mode distribution of inelastically scattered electrons.
\end{abstract}

\maketitle

\section{Introduction}

Electrons scattered by an interaction with matter, such as from individual atoms, molecules, or materials, acquire a spectrum of energies, linear momenta, and spin polarizations. Information about the event is encoded in these various degrees of freedom by the electron's wavefunction. Recently, several groups demonstrated control of the orbital angular momentum (OAM) of freedom of free electrons \cite{uchida_generation_2010, verbeeck_production_2010,mcmorran_electron_2011}. Myriad techniques for generating electron OAM states now exist, including material and magnetic spiral phase plates \cite{uchida_generation_2010, blackburn_vortex_2014, beche_magnetic_2014, beche_focused_2016}, phase \cite{harvey_efficient_2014,grillo_highly_2014,shiloh_sculpturing_2014} and amplitude \cite{verbeeck_production_2010, saitoh_production_2012} diffraction gratings, and mode conversion \cite{schattschneider_novel_2012}. Exchange of OAM between a target specimen and a fast electron could provide information about the structural chirality \cite{asenjo-garcia_dichroism_2014, harvey_demonstration_2015} and out-of-plane magnetization of the target \cite{idrobo_vortex_2011, lloyd_quantized_2012, schattschneider_mapping_2012}. In these applications of electron orbital angular momentum, the electron beam can scatter to many different final orbital angular momentum states. These applications can therefore offer more information with measurement of the final orbital angular momentum distribution. No orbital angular momentum measurement techniques exist that can efficiently and quantitatively measure the final orbital angular momentum distribution.

\begin{figure}
  \includegraphics[width = 0.8\columnwidth]{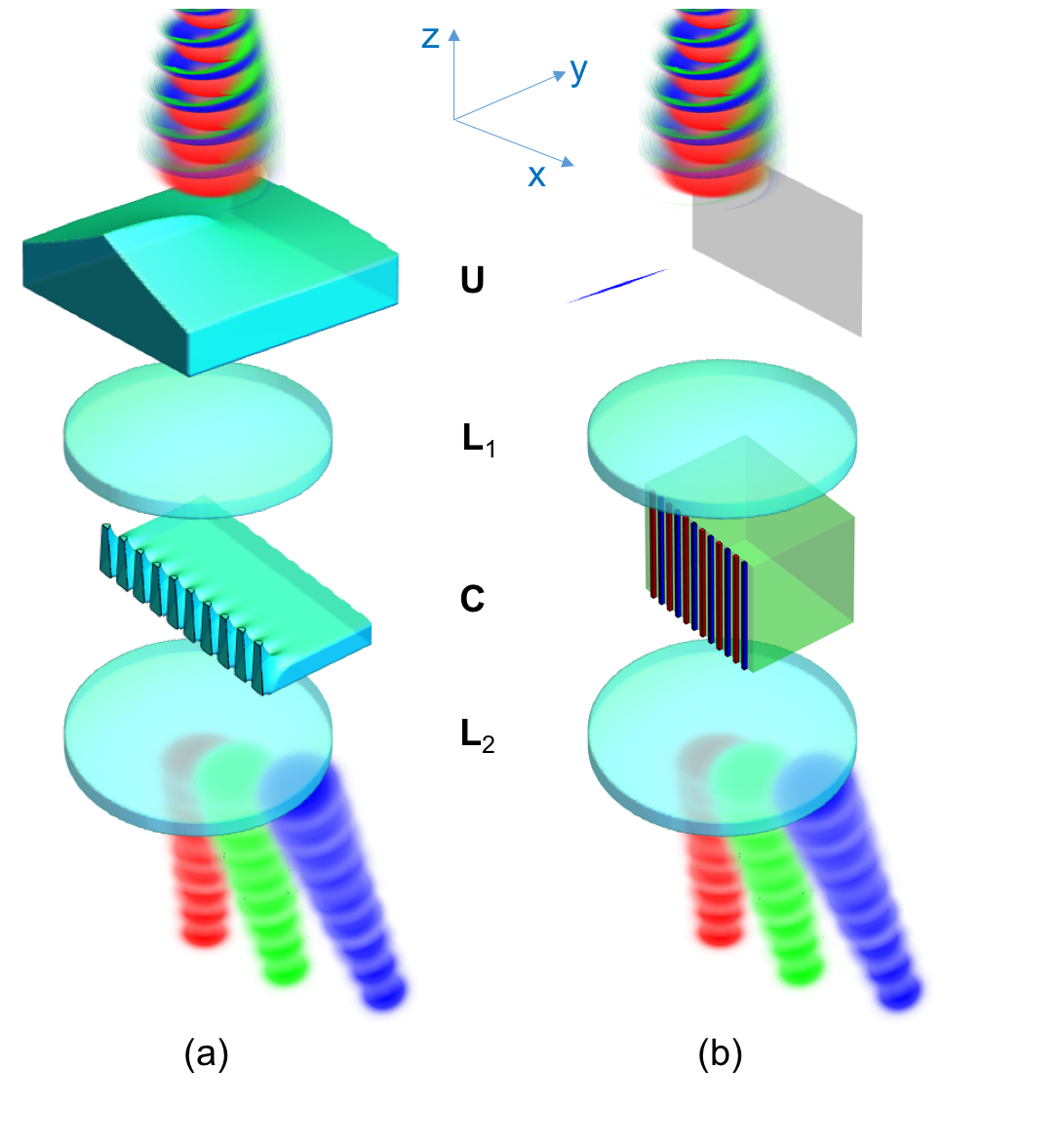}
  \caption{Schematic of the optical arrangement of OAM-sorting devices for (a) light and (b) electrons. Different OAM states are shown in different colors. Mixed OAM states are incident on the top of both systems each of which consists of four elements. A phase unwrapper element U in the front focal plane of a lens L1 is followed by a phase corrector element C in the back focal place of L1. For electrons, the proposed element U is a charged needle or knife edge, and the corrector element C is an array of electrodes with alternating bias. Immediately after the corrector element C, different OAM components are separated in momentum space. At the bottom of each device, a Fourier-transforming lens L2 separates OAM components into different spots in real space at the output.}
  \label{fig:OAMsorter}
\end{figure}

In 2010, Berkhout et al. \cite{berkhout_efficient_2010} demonstrated a new method to efficiently sort OAM states of light using four refractive optical elements. The apparatus transforms an azimuthal phase at the input into a linear phase at the output, such that OAM components at the input are mapped into separate linear momentum states at the output. This ability to measure superpositions and mixed states of optical OAM enables parallel orbital angular momentum measurement. The apparatus has been rapidly employed for a range of optical applications in both fundamental research \cite{lavery_refractive_2012,malik_direct_2014}, quantum information \cite{mirhosseini_high-dimensional_2015}, and communications \cite{yan_high-capacity_2014,huang_mode_2015}. As shown in the left side of Fig. \ref{fig:OAMsorter}, the apparatus consists of a phase unwrapper, a lens, and a phase corrector.   As shown in the left side of Fig. \ref{fig:OAMsorter}, the apparatus is based on two custom-made non-spherical refractive optical components, the phase unwrapper U and the phase corrector C, with two lenses L1 and L2 used to the Fourier transform the output of each. 

\begin{figure}
  \subfloat[]{\label{subfig:U_phase_profile}
  \includegraphics[width = 0.5\columnwidth]{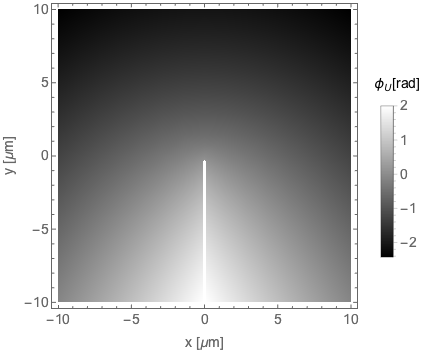}}
  \subfloat[]{\label{subfig:C_phase_profile}
  \includegraphics[width = 0.5\columnwidth]{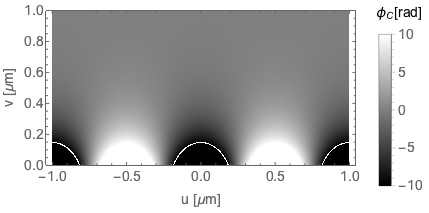}}
  \caption{Phase profiles of the (a) unwrapper element U described in Equation \ref{eq:phi1} and (b) the corrector element C described in Equation \ref{eq:phi2}. Both plots use parameters expressed in Table \ref{table:params}.}
  \label{fig:sorterphaseplates}
\end{figure}

The first optical element (element U in Fig. \ref{fig:OAMsorter}) is a log-polar transformerÊ\cite{hossack_coordinate_1987} that transforms a set of concentric rings at the input plane into a set of parallel lines at the back focal plane of the lens. The phase profile of this unwrapper element is described by Eq. 1 inÊ\cite{berkhout_efficient_2010}:

\begin{equation}
	\varphi_u(x,y) = \frac{d}{\lambda f}\left[x \arctan{\left(\frac{x}{y}\right)}+y-y \ln{\left(\frac{\sqrt{x^2+y^2}}{b}\right)}  \right],
    \label{eq:phi1}
\end{equation}

\noindent where here we adopt a coordinate system rotated from \cite{berkhout_efficient_2010}, $d$ is a lengthscale associated with the output optical distribution, and $f$ is the focal length of the first lens L1 following the phase unwrapper U. A plot of the phase distribution for this lens are shown in Fig. \ref{fig:sorterphaseplates}. 

\section{Electrostatic OAM Unwrapper for Electrons}
To imprint the phase profile described in Eq. \ref{eq:phi1} onto an electron wavefunction, one could use either refractive or diffractive wavefront-shaping techniques. In light optics, there are established methods for fabricating custom phase plates out of transparent material such as glass. However, while thin film phase plates for electrons are possible \cite{beche_focused_2016}, they contaminate easily and are difficult to fabricate. Finally, no material is sufficiently electron-transparent to imprint the large phases required for sorting OAM.
Arbitrary electron phase profiles can be imprinted holographically using nanofabricated diffractive optics \cite{harvey_efficient_2014, grillo_highly_2014}. However, the smaller but still significant inelastic scattering in the material, the small diffraction angles, low diffraction efficiency, and finite size of the diffractive structures make the use of such holograms for an OAM mode sorter impractical. 


%

Instead, a relatively simple electrostatic phase plate consisting of a charged needle and a conductive plate can be used to imprint a phase equivalent to Eq. \ref{eq:phi1} onto a charged particle wave. The phase that the tip of a charged needle imparts to an electron has been studied previously by several different groups \cite{matteucci_electron_1992, cumings_electron_2002, beleggia_towards_2014}. Matteucci \emph{et al.} \cite{matteucci_electron_1992} calculated this analytically by first considering the electrostatic potential $V(\mathbf{r})$ around an infinitesimally thin wire of finite length and uniform charge density placed a distance $h$ away from a flat conducting plate. The spatially varying phase shift a potential $V(\mathbf{r})$ imparts to an electron plane wave of energy $E$ and relativistically-corrected wavelength $\lambda$ traveling in the $+z$ direction can be calculated by the integral 

\begin{equation}
  \varphi(\mathbf{r}) = C_E \int^{\infty}_{-\infty} V(\mathbf{r}) dz,.
  \label{eq:vp1a}
\end{equation}

\noindent where $C_E$ is a constant that depends only on the energy of the beam \cite{mccartney_electron_2007} ($C_E = 6.53$ mrad V$^{-1}$ nm$^{-1}$ for 300 keV electrons). 

In Appendix A, we adapt Matteucci \emph{et al.}'s result (Equation 4 in \cite{matteucci_electron_1992}) for the purpose of imprinting Eq. \ref{eq:phi1}.  We show that if the electron beam is localized around the needle tip nearest the plate electrode, and the length of the needle and its separation from the plate are sufficiently large, this arrangement imprints the appropriate unwrapping phase for sorting electron OAM:

\begin{equation}
	\varphi_{\textrm{tip}}(x,y) = \frac{Q C_E}{2 \pi \epsilon_0 L}\left[ x \arctan{\left(\frac{x}{y}\right)} + y \ln \left(\frac{\sqrt{x^2+y^2}}{L}\right)\right] + \varphi_0,
    \label{eq:vpend}
\end{equation}
\noindent where $L$ is the length of the needle and $\varphi_0$ is a uniform phase common to all paths.

We also note that an extended knife edge electrode could potentially be used instead of a charged needle. The 2D electrostatic potential of a semi-infinite plane of charge with it's edge along the $z$-axis has the same functional form as the desired unwrapper phase $\varphi_u(x,y)$ (see Chapter 7 in \cite{feynman_chapter_1964}). Thus, a knife-edge electrode aligned with the optical axis could provide an alternative design to the needle, if the length were long enough such that phase introduced near the beginning and end of the electrode were negligible.

\section{Electrostatic Phase Corrector for Electrons}
The phase unwrapper element is followed by a conventional electron lens system (L1). Simulations of the electron wave function in the back focal plane of this intermediate lens show that there are large variations in the phase due to the unwrapping operation. These phase variations must be removed by a second optical element to reveal the subtler OAM-dependent differences. This phase corrector (Fig. \ref{fig:sorterphaseplates}) is described by the following phase profile:

\begin{equation}
  \varphi_c(u,v) = \frac{b d}{\lambda f}\exp{\left(-\frac{2 \pi u}{d}\right)}\cos\left(-\frac{2 \pi v}{d}\right),
    \label{eq:phi2}
\end{equation}
\noindent
where the lengthscale $b$ describes the separation of OAM components at the output and $f$ is also the focal length of the second Fourier-transforming lens L2.

\begin{figure}
  \subfloat[]{\label{subfig:needle_cartoon}
  \includegraphics[width=0.5\columnwidth]{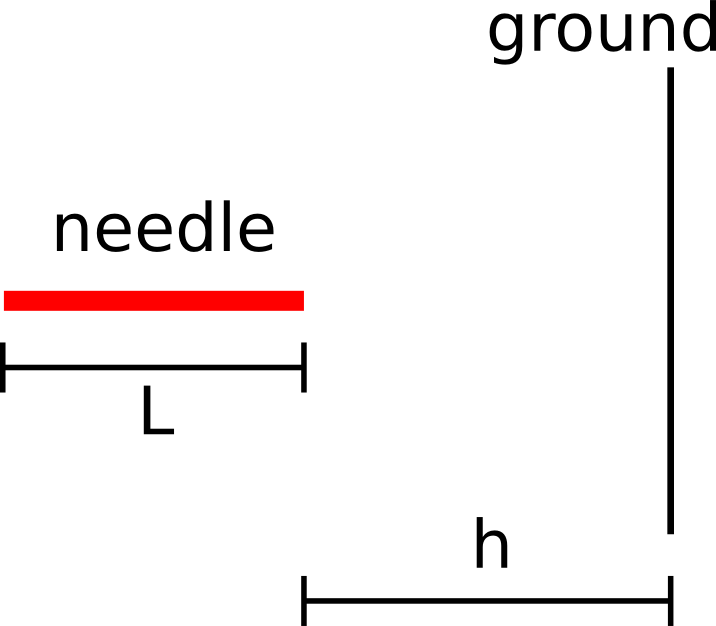}} \\
  \subfloat[]{\label{subfig:corrector_cartoon}
  \includegraphics[width=0.5\columnwidth]{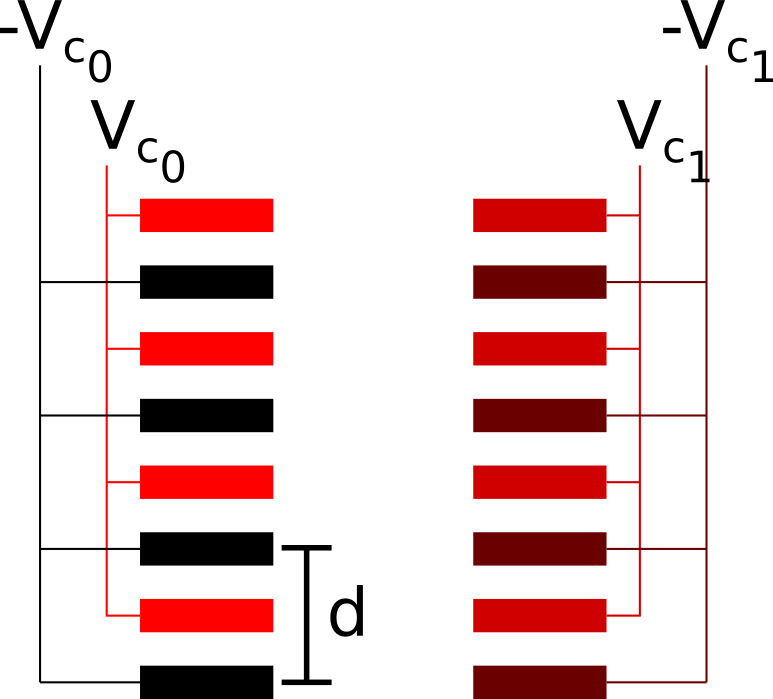}}
  \caption{(a) Top-view cartoon of charged needle and ground plate used to produce the unwrapper phase. Electrons passing into the device near the right end of the needle will acquire the phase described by Eq. \ref{eq:vpend}. (b) Top-view cartoon of example electrodes that could be used to produce the corrector phase, \ref{eq:phi2}. Alternating very high ($V_{c_0}$) and very low ($-V_{c_0}$) voltages at the boundary at $u = 0$ produce a sinusoidal potential in the $v$-direction. Alternating weakly high ($V_{c_1}$) and weakly low ($-V_{c_1}$) voltages at the boundary at $u = u_{1}$ produce an exponential decay in the $u$-direction. The electrodes at $u = u_{1}$ aren't physically necessary, as we show in Fig. \ref{subfig:corrector:x-y_slice_flat} \label{fig:corrector_cartoon}}
\end{figure}
Electrostatic elements can also be employed to imprint this corrector phase. As the phase distribution is a solution to Laplace's equation in 2D, i.e. $\nabla^2 \varphi_c(u,v) = 0$, we see that an electrostatic potential in 2D can take this form. We can approximate the 2D potential solution in 3D with a potential that varies slowly in $z$. Specifically, we can apply $\varphi_c(u,v)$ to an electron with a set of alternating electrodes, as shown in Fig. \ref{fig:corrector_cartoon}. As long as the longitudinal height $D$ of the electrodes is much longer than the period $d$ (see Appendix \ref{sect:corrector_depth}), and the thin grating condition, $\lambda D \ll d^2$ is satisfied, the variation of the potential in the longitudinal direction is negligible over the depth. The corrector phase can be written as
\begin{equation}
  \varphi_c(u,v) = C_E D V_{c_0} \exp{\left(\frac{-2 \pi u}{d}\right)}\cos\left(\frac{-2 \pi v}{d}\right). \label{eq:corrector_phase}
\end{equation}
\noindent
We see that we get the appropriate $\varphi_c$ (Eq. \ref{eq:phi2}) if $C_EDV_{c_0} = \frac{b d}{\lambda f}$ and $V_{c_1} = V_{c_0} \exp\left(-\frac{2 \pi u_1}{d}\right)$. Further analysis (Appendix \ref{sect:corrector_depth}) shows that it could be practical to replace the reference electrodes (held at $\pm V_{c_1}$ in Fig. \ref{fig:corrector_cartoon}b) by a single plate, or even remove this reference surface altogether.

The final spacing between modes is
\begin{equation}
  \Delta t = \frac{\lambda f}{d}.
\end{equation}
Lavery {\it et al.} separated orbital angular momentum states of light with a wavelength of $\lambda = \unit[632.8]{\textrm{nm}}$, lens focal length $f = \unit[300]{\textrm{mm}}$, a corrector period $d = \unit[8]{\textrm{mm}}$ and therefore an unmagnified separation of $\Delta t = \unit[23.73]{\mu \textrm{m}}$ \cite{lavery_refractive_2012}. As preparation of a collimated photon orbital angular momentum state with a waist on the order of $\unit[10]{\mu\textrm{m}}$ is straightforward, this separation is sufficient.

The orders of magnitude of these parameters are wildly different for electrons, but good separation is similarly straightforward. With a needle length of $L \sim \unit[50]{\mu\textrm{m}}$, an incident beam waist on the order of $\unit[1]{\mu\textrm{m}}$ is physically reasonable. Separation on the order of $\Delta t = \unit[0.32]{\mu \textrm{m}}$ can be achieved in a transmission electron microscope at $\unit[300]{\textrm{kV}}$ with $\lambda \sim \unit[1.97]{\textrm{pm}}$ and a corrector period of $d \sim \unit[1]{\mu\textrm{m}}$ if the focal length of the lens between the needle and corrector, L1, is $f \sim \unit[100]{\textrm{cm}}$. Several lenses with focal lengths in the $\unit[1]{\textrm{cm}}$ to $\unit[10]{\textrm{cm}}$ range can be combined to more practically produce a 1 meter focal length over a much shorter distance.

\begin{table}[hb] 
\caption{} \label{table:params}
  \centering
  \begin{tabular}{ >{$}l<{$} | >{$}r<{$} }
    \textrm{Sorter Parameter} &	\textrm{Magnitude} \\
  \hline
  \lambda & \unit[1.97]{\textrm{pm}} \\
  f & \unit[100]{\textrm{cm}}  \\
  d & \unit[1]{\mu \textrm{m}} \\
  b = L & \unit[50]{\mu \textrm{m}} \\
  V_{c_0}D & \unit[38]{\textrm{V}\cdot \mu \textrm{m}} \\
  Q/L & \unit[42]{\mu \textrm{C}/\mu \textrm{m}} \\
  \end{tabular}
\end{table}

To review, the parameters of this arrangement are: (a) the charge $Q$ added to the needle-based unwrapper phase plate, (b) the length of the needle $L$,
 (c) the voltage $V_{c_0}$ applied to the corrector electrodes, (d) the spatial periodicity $d$ of the corrector electrodes, and (e) the focal length of the lenses $f$.

\begin{widetext}

\begin{figure}
  \subfloat[]{\label{subfig:superposition_input33}
  \includegraphics[width=0.3\columnwidth]{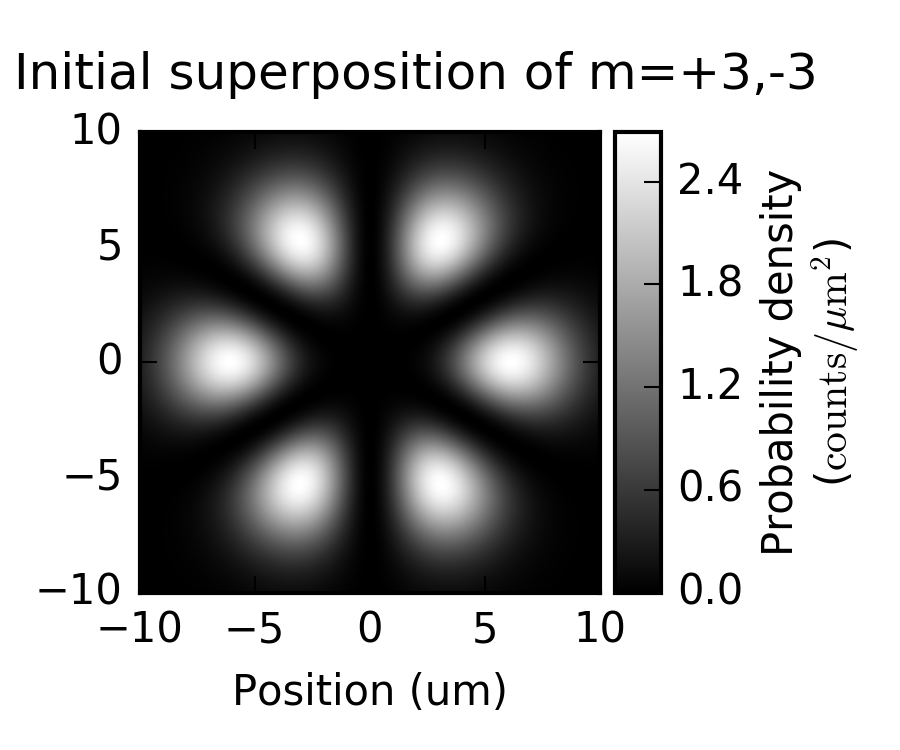} } 
  \subfloat[]{\label{subfig:superposition_input52}
  \includegraphics[width=0.3\columnwidth]{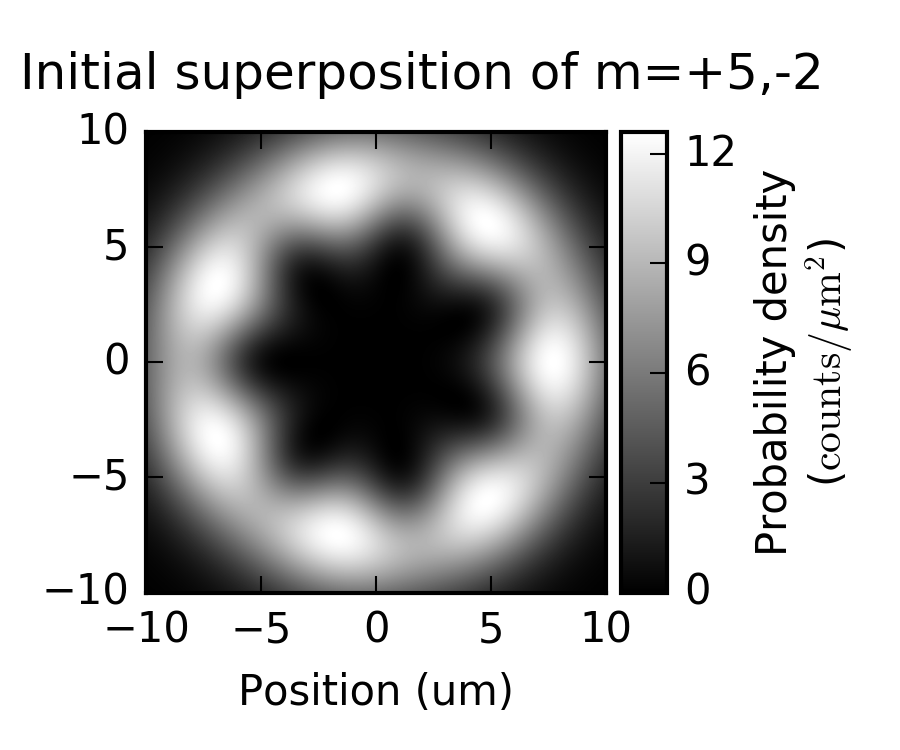} } 
  \subfloat[]{\label{subfig:superposition_input30}
  \includegraphics[width=0.3\columnwidth]{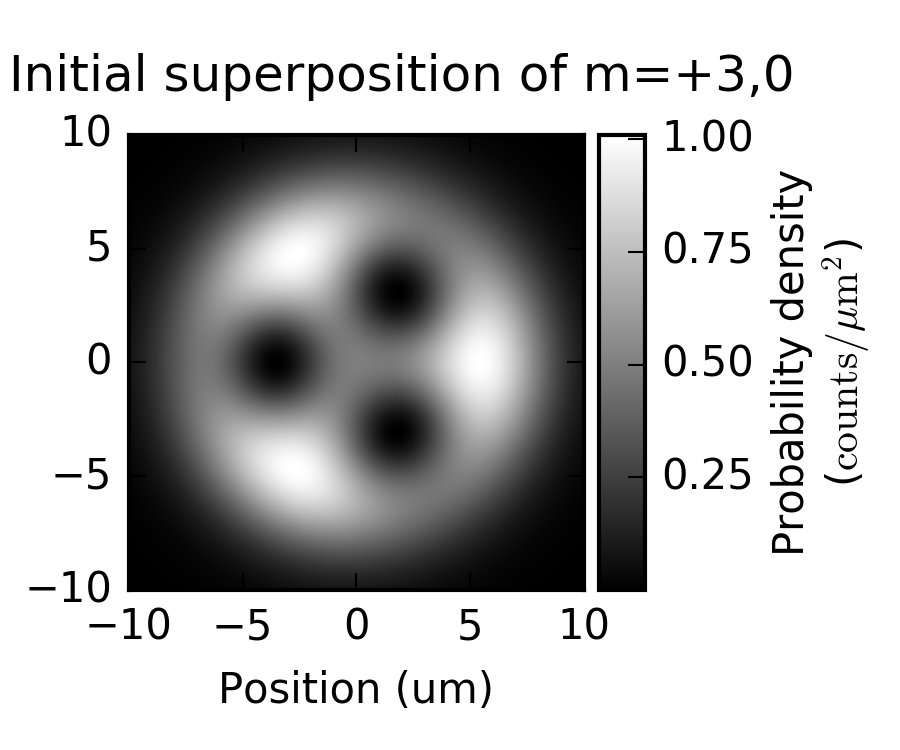} }  \\
  \subfloat[]{\label{subfig:superposition_output33}
  \includegraphics[width=0.3\columnwidth]{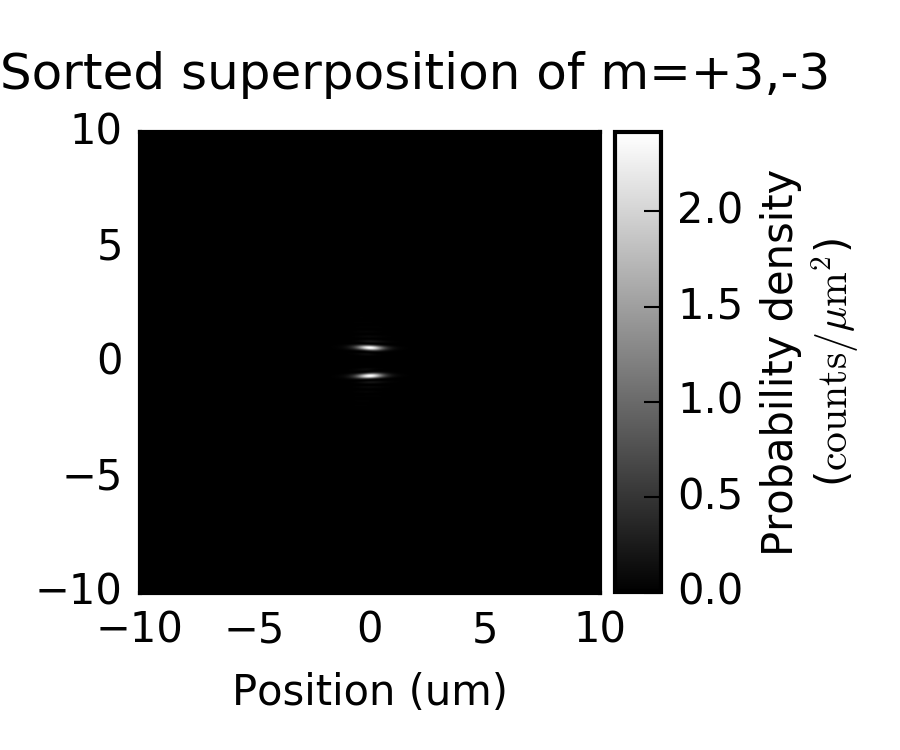} } 
  \subfloat[]{\label{subfig:superposition_output52}
  \includegraphics[width=0.3\columnwidth]{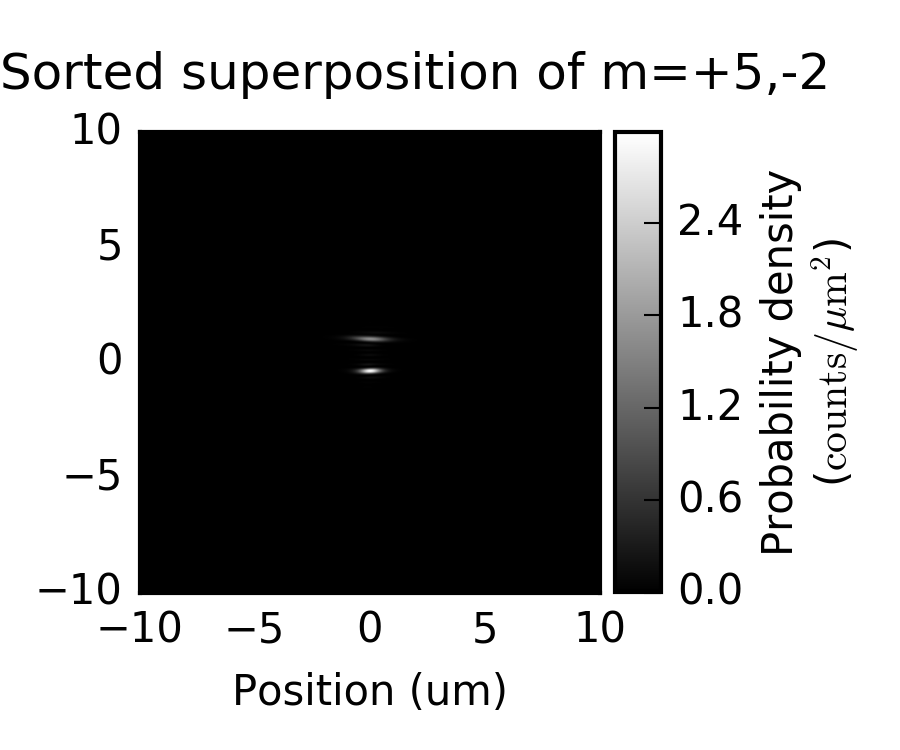} }
  \subfloat[]{\label{subfig:superposition_output30}
  \includegraphics[width=0.3\columnwidth]{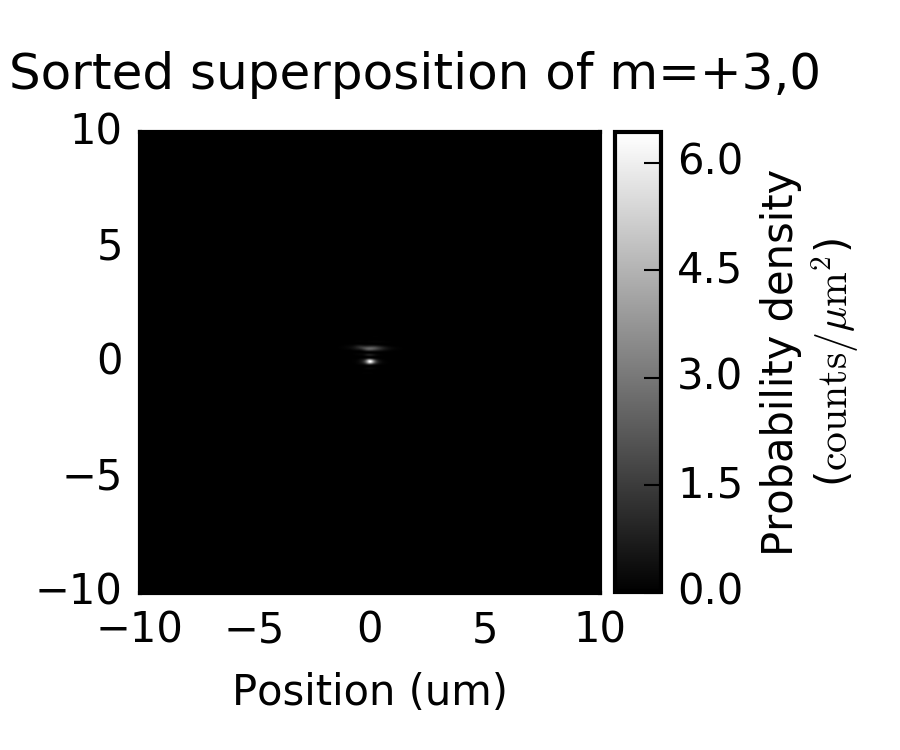} } \\
  \caption{The simulated (a,b,c) input and (d,e,f) output of the proposed electron OAM sorter using parameters shown in Table \ref{table:params}. Input states are superpositions of Laguerre-Gaussian modes with and a $\unit[5]{\mu \textrm{m}}$ beam waist and (a) $m=+ 3$ and $m=-3$, (b) $m=+5$ and $m = -2$, and (c) $m=3$ and $m=0$. Each electron OAM component at the input gets mapped onto a separate region in space at the output, which is viewed directly using TEM imaging optics. In this way, a spectrum of electron OAM states can be efficiently recorded in parallel.  \label{fig:sorteroutput}}
\end{figure}

\begin{figure}[h]
  \subfloat[]{\label{subfig:corrected_plus_ampphase}
  \includegraphics[height=2.5in]{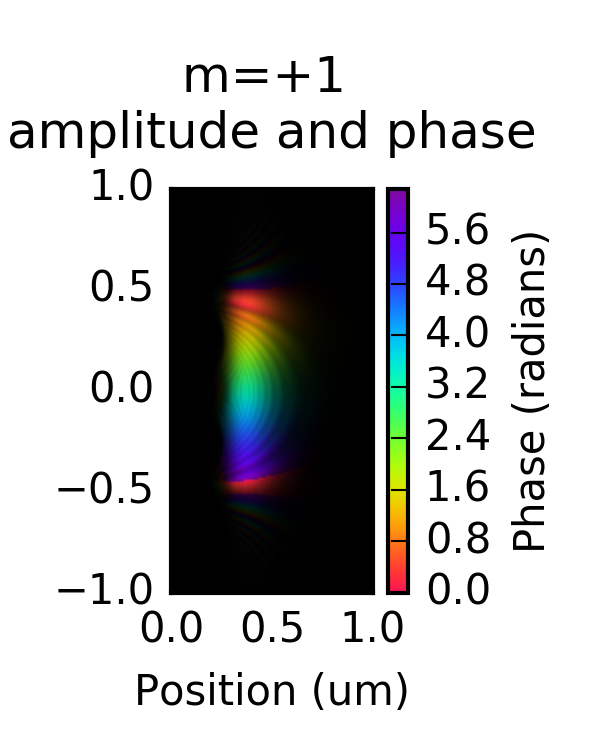} } 
  \subfloat[]{\label{subfig:plus_output}
  \includegraphics[height=2.5in]{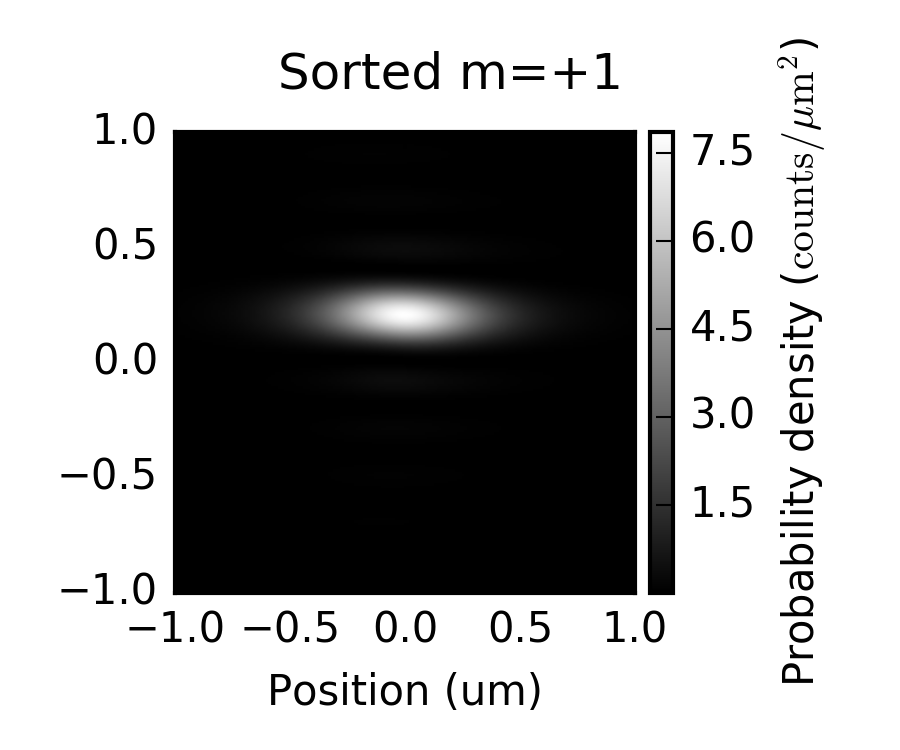} } \\
  \subfloat[]{\label{subfig:corrected_minus_ampphase}
  \includegraphics[height=2.5in]{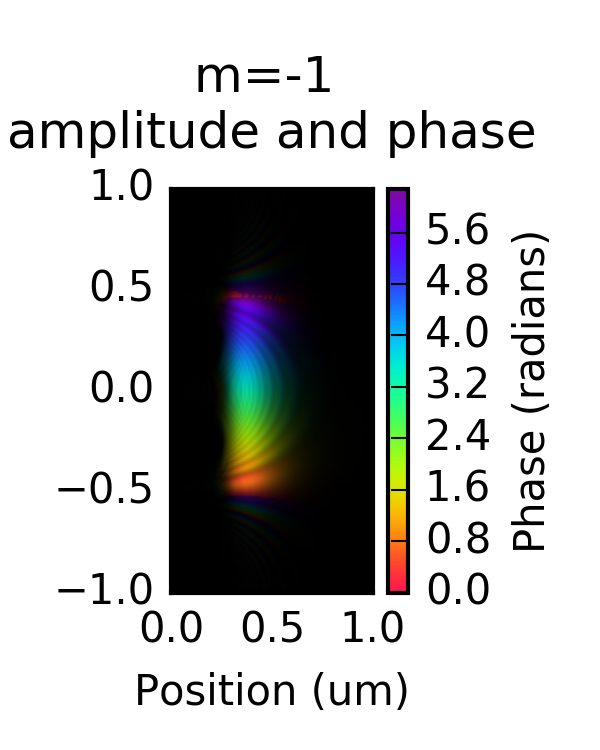} } 
  \subfloat[]{\label{subfig:minus_output}
  \includegraphics[height=2.5in]{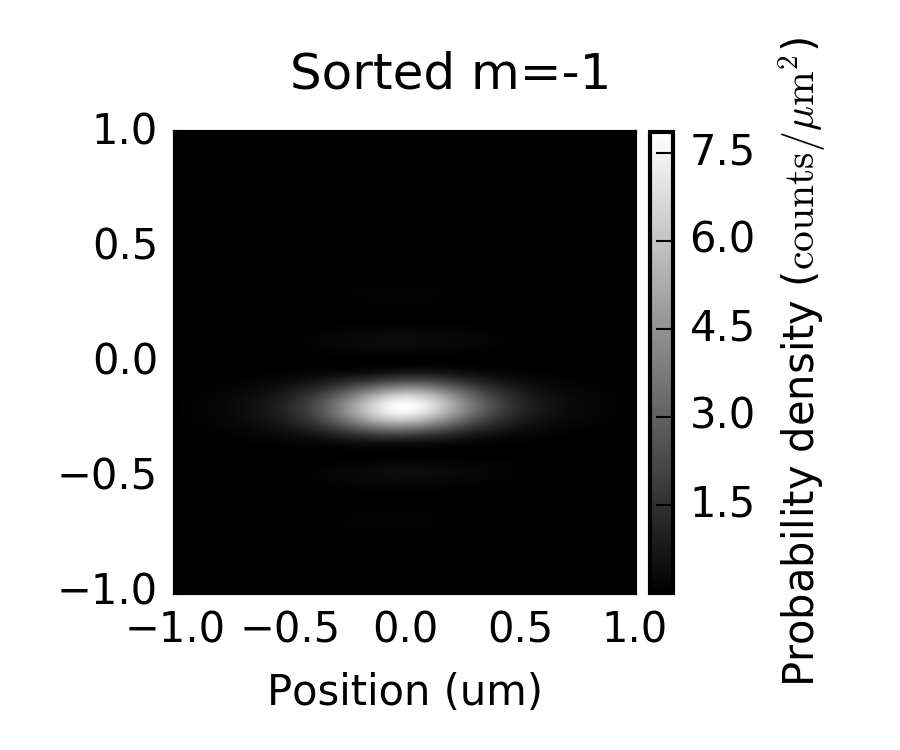} } \\
  \caption{Illustration of the action of the device: the unwrapper and corrector produce plane wave-like beams in the corrector plane (a,c) which correspond to deflected spots in the output plane (b,d). Sorter simulated with parameters shown in Table \ref{table:params}.  \label{fig:sorter_action}}
\end{figure}

\begin{figure}[h]
  \subfloat[]{\label{subfig:initial_globs}
  \includegraphics[height=2.5in]{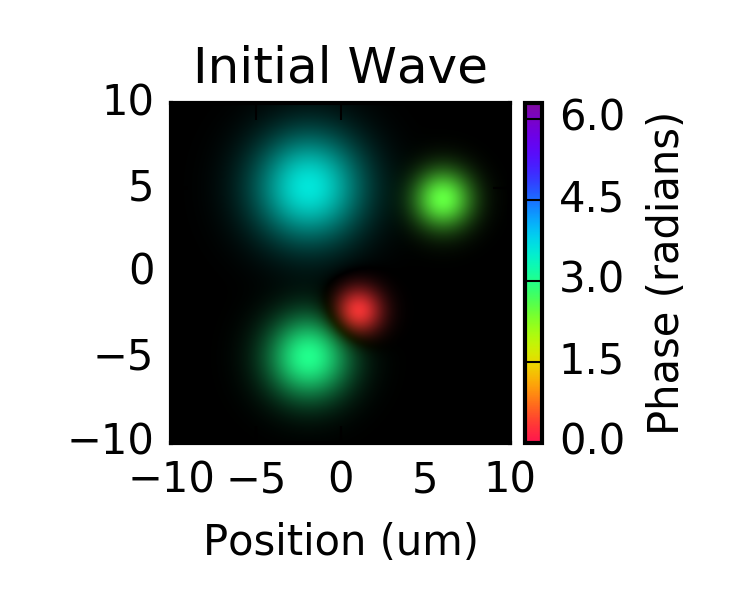}}
  \subfloat[]{\label{subfig:calculated_distribution}
  \includegraphics[height=2.5in]{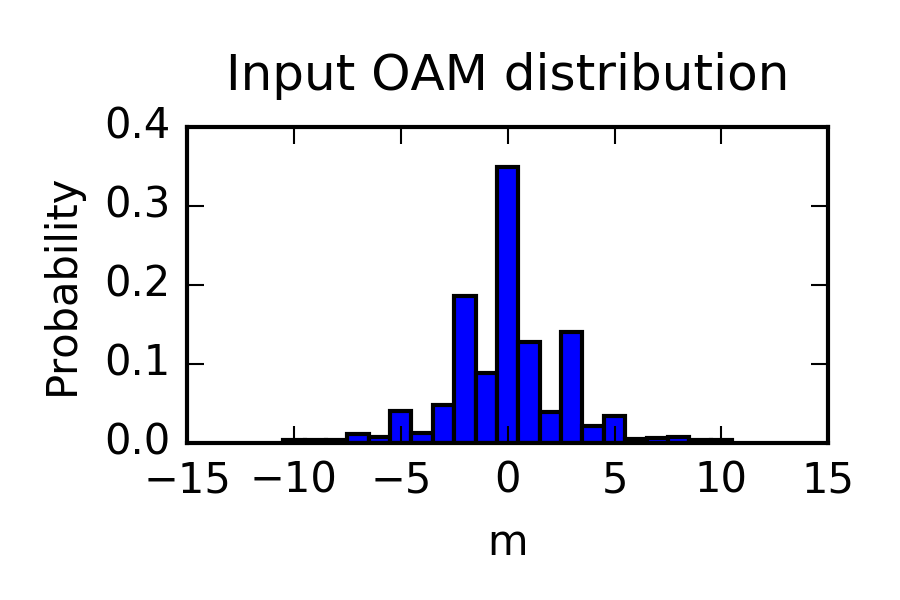}}\\
  \subfloat[]{\label{subfig:sorted_globs}
  \includegraphics[height=2.5in]{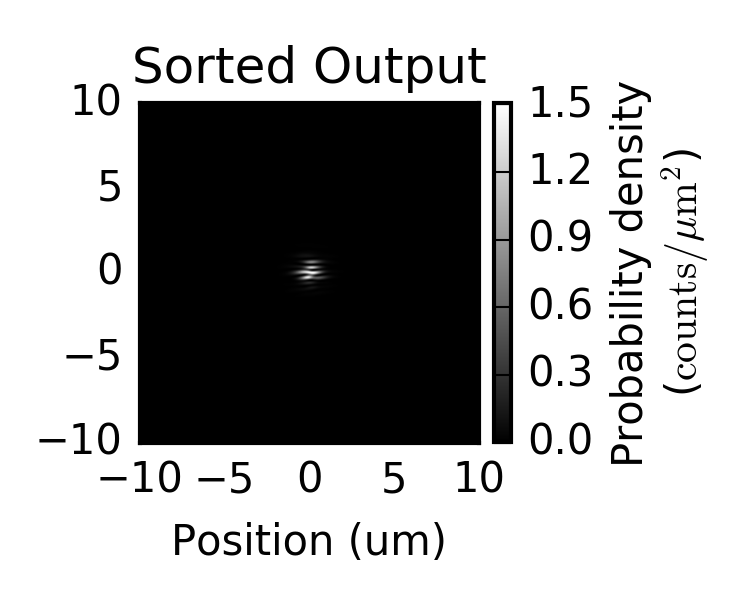} } 
  \subfloat[]{\label{subfig:sorter_distribution}
  \includegraphics[height=2.5in]{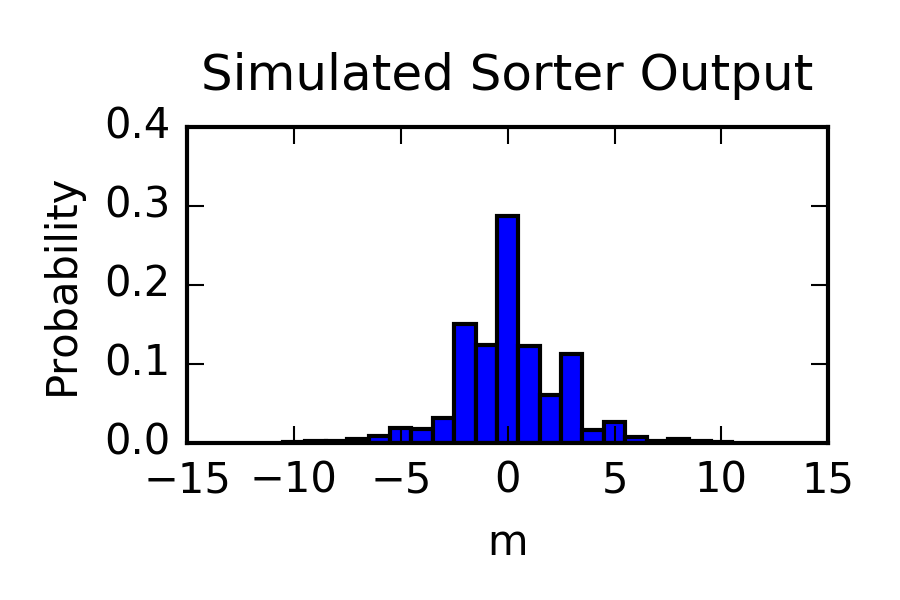}} \\
  \caption{(a) Initial random wavefunction with non-trivial orbital angular momentum distribution; amplitude is shown as brightness and phase is shown as hue; (b) calculated orbital angular angular momentum distribution; (c) probability density of the random wavefunction passed through the sorter; (d) orbital angular momentum distribution calculated by binning the output of the sorter. \label{fig:globs}}
\end{figure}

\end{widetext}


\section{Conclusion}
Knowledge of interactions in which a free electron exchanges OAM with a specimen can lead to insights into the properties of the object. However, many attempts by several groups to observe OAM transfer between a prepared focused electron with OAM and an atom have so far been unsuccessful, due to the fact that electrons are scattered into a superposition of orbital states. Here we described an electron-optical analog of the OAM sorter developed for photons. This device can non-destructively disperse the spectrum of electron OAM, providing a way to measure the OAM distribution of electrons scattered or ejected from atoms, molecules, and larger collections of matter. Thus, this could provide a completely new form of spectroscopy that can be used to probe the asymmetric structure of matter, atomic and molecular polarizations, and chiral interactions. 

\begin{acknowledgements} 
  This work at UO was partially supported by the U.S. Department of Energy, Office of Science, Basic Energy Sciences, under Award \#DE-SC0010466. We thank Martin Linck of Corrected Electron Optical Systems GmbH for order-of-magnitude estimates of lens focal lengths that were helpful in assessing practical relevance.
\end{acknowledgements}

\bibliographystyle{apsrev4-1}
\bibliography{eOAM_sorter}{}

\begin{thebibliography}{28}%
\makeatletter
\providecommand \@ifxundefined [1]{%
 \@ifx{#1\undefined}
}%
\providecommand \@ifnum [1]{%
 \ifnum #1\expandafter \@firstoftwo
 \else \expandafter \@secondoftwo
 \fi
}%
\providecommand \@ifx [1]{%
 \ifx #1\expandafter \@firstoftwo
 \else \expandafter \@secondoftwo
 \fi
}%
\providecommand \natexlab [1]{#1}%
\providecommand \enquote  [1]{``#1''}%
\providecommand \bibnamefont  [1]{#1}%
\providecommand \bibfnamefont [1]{#1}%
\providecommand \citenamefont [1]{#1}%
\providecommand \href@noop [0]{\@secondoftwo}%
\providecommand \href [0]{\begingroup \@sanitize@url \@href}%
\providecommand \@href[1]{\@@startlink{#1}\@@href}%
\providecommand \@@href[1]{\endgroup#1\@@endlink}%
\providecommand \@sanitize@url [0]{\catcode `\\12\catcode `\$12\catcode
  `\&12\catcode `\#12\catcode `\^12\catcode `\_12\catcode `\%12\relax}%
\providecommand \@@startlink[1]{}%
\providecommand \@@endlink[0]{}%
\providecommand \url  [0]{\begingroup\@sanitize@url \@url }%
\providecommand \@url [1]{\endgroup\@href {#1}{\urlprefix }}%
\providecommand \urlprefix  [0]{URL }%
\providecommand \Eprint [0]{\href }%
\providecommand \doibase [0]{http://dx.doi.org/}%
\providecommand \selectlanguage [0]{\@gobble}%
\providecommand \bibinfo  [0]{\@secondoftwo}%
\providecommand \bibfield  [0]{\@secondoftwo}%
\providecommand \translation [1]{[#1]}%
\providecommand \BibitemOpen [0]{}%
\providecommand \bibitemStop [0]{}%
\providecommand \bibitemNoStop [0]{.\EOS\space}%
\providecommand \EOS [0]{\spacefactor3000\relax}%
\providecommand \BibitemShut  [1]{\csname bibitem#1\endcsname}%
\let\auto@bib@innerbib\@empty
\bibitem [{\citenamefont {Uchida}\ and\ \citenamefont
  {Tonomura}(2010)}]{uchida_generation_2010}%
  \BibitemOpen
  \bibfield  {author} {\bibinfo {author} {\bibfnamefont {M.}~\bibnamefont
  {Uchida}}\ and\ \bibinfo {author} {\bibfnamefont {A.}~\bibnamefont
  {Tonomura}},\ }\href {\doibase 10.1038/nature08904} {\bibfield  {journal}
  {\bibinfo  {journal} {Nature}\ }\textbf {\bibinfo {volume} {464}},\ \bibinfo
  {pages} {737} (\bibinfo {year} {2010})},\ \bibinfo {note} {00172 Cited by
  0096}\BibitemShut {NoStop}%
\bibitem [{\citenamefont {Verbeeck}\ \emph {et~al.}(2010)\citenamefont
  {Verbeeck}, \citenamefont {Tian},\ and\ \citenamefont
  {Schattschneider}}]{verbeeck_production_2010}%
  \BibitemOpen
  \bibfield  {author} {\bibinfo {author} {\bibfnamefont {J.}~\bibnamefont
  {Verbeeck}}, \bibinfo {author} {\bibfnamefont {H.}~\bibnamefont {Tian}}, \
  and\ \bibinfo {author} {\bibfnamefont {P.}~\bibnamefont {Schattschneider}},\
  }\href {\doibase 10.1038/nature09366} {\bibfield  {journal} {\bibinfo
  {journal} {Nature}\ }\textbf {\bibinfo {volume} {467}},\ \bibinfo {pages}
  {301} (\bibinfo {year} {2010})},\ \bibinfo {note} {00221 Cited by
  0114}\BibitemShut {NoStop}%
\bibitem [{\citenamefont {McMorran}\ \emph {et~al.}(2011)\citenamefont
  {McMorran}, \citenamefont {Agrawal}, \citenamefont {Anderson}, \citenamefont
  {Herzing}, \citenamefont {Lezec}, \citenamefont {McClelland},\ and\
  \citenamefont {Unguris}}]{mcmorran_electron_2011}%
  \BibitemOpen
  \bibfield  {author} {\bibinfo {author} {\bibfnamefont {B.~J.}\ \bibnamefont
  {McMorran}}, \bibinfo {author} {\bibfnamefont {A.}~\bibnamefont {Agrawal}},
  \bibinfo {author} {\bibfnamefont {I.~M.}\ \bibnamefont {Anderson}}, \bibinfo
  {author} {\bibfnamefont {A.~A.}\ \bibnamefont {Herzing}}, \bibinfo {author}
  {\bibfnamefont {H.~J.}\ \bibnamefont {Lezec}}, \bibinfo {author}
  {\bibfnamefont {J.~J.}\ \bibnamefont {McClelland}}, \ and\ \bibinfo {author}
  {\bibfnamefont {J.}~\bibnamefont {Unguris}},\ }\href {\doibase
  10.1126/science.1198804} {\bibfield  {journal} {\bibinfo  {journal}
  {Science}\ }\textbf {\bibinfo {volume} {331}},\ \bibinfo {pages} {192 }
  (\bibinfo {year} {2011})}\BibitemShut {NoStop}%
\bibitem [{\citenamefont {Blackburn}\ and\ \citenamefont
  {Loudon}(2014)}]{blackburn_vortex_2014}%
  \BibitemOpen
  \bibfield  {author} {\bibinfo {author} {\bibfnamefont {A.~M.}\ \bibnamefont
  {Blackburn}}\ and\ \bibinfo {author} {\bibfnamefont {J.~C.}\ \bibnamefont
  {Loudon}},\ }\href {\doibase 10.1016/j.ultramic.2013.08.009} {\bibfield
  {journal} {\bibinfo  {journal} {Ultramicroscopy}\ }\textbf {\bibinfo {volume}
  {136}},\ \bibinfo {pages} {127} (\bibinfo {year} {2014})},\ \bibinfo {note}
  {00013}\BibitemShut {NoStop}%
\bibitem [{\citenamefont {B{\'e}ch{\'e}}\ \emph {et~al.}(2014)\citenamefont
  {B{\'e}ch{\'e}}, \citenamefont {Van~Boxem}, \citenamefont {Van~Tendeloo},\
  and\ \citenamefont {Verbeeck}}]{beche_magnetic_2014}%
  \BibitemOpen
  \bibfield  {author} {\bibinfo {author} {\bibfnamefont {A.}~\bibnamefont
  {B{\'e}ch{\'e}}}, \bibinfo {author} {\bibfnamefont {R.}~\bibnamefont
  {Van~Boxem}}, \bibinfo {author} {\bibfnamefont {G.}~\bibnamefont
  {Van~Tendeloo}}, \ and\ \bibinfo {author} {\bibfnamefont {J.}~\bibnamefont
  {Verbeeck}},\ }\href {\doibase 10.1038/nphys2816} {\bibfield  {journal}
  {\bibinfo  {journal} {Nature Physics}\ }\textbf {\bibinfo {volume} {10}},\
  \bibinfo {pages} {26} (\bibinfo {year} {2014})},\ \bibinfo {note}
  {00034}\BibitemShut {NoStop}%
\bibitem [{\citenamefont {B{\'e}ch{\'e}}\ \emph {et~al.}(2016)\citenamefont
  {B{\'e}ch{\'e}}, \citenamefont {Winkler}, \citenamefont {Plank},
  \citenamefont {Hofer},\ and\ \citenamefont {Verbeeck}}]{beche_focused_2016}%
  \BibitemOpen
  \bibfield  {author} {\bibinfo {author} {\bibfnamefont {A.}~\bibnamefont
  {B{\'e}ch{\'e}}}, \bibinfo {author} {\bibfnamefont {R.}~\bibnamefont
  {Winkler}}, \bibinfo {author} {\bibfnamefont {H.}~\bibnamefont {Plank}},
  \bibinfo {author} {\bibfnamefont {F.}~\bibnamefont {Hofer}}, \ and\ \bibinfo
  {author} {\bibfnamefont {J.}~\bibnamefont {Verbeeck}},\ }\href {\doibase
  10.1016/j.micron.2015.07.011} {\bibfield  {journal} {\bibinfo  {journal}
  {Micron}\ }\textbf {\bibinfo {volume} {80}},\ \bibinfo {pages} {34} (\bibinfo
  {year} {2016})},\ \bibinfo {note} {00002}\BibitemShut {NoStop}%
\bibitem [{\citenamefont {Harvey}\ \emph {et~al.}(2014)\citenamefont {Harvey},
  \citenamefont {Pierce}, \citenamefont {Agrawal}, \citenamefont {Ercius},
  \citenamefont {Linck},\ and\ \citenamefont
  {McMorran}}]{harvey_efficient_2014}%
  \BibitemOpen
  \bibfield  {author} {\bibinfo {author} {\bibfnamefont {T.~R.}\ \bibnamefont
  {Harvey}}, \bibinfo {author} {\bibfnamefont {J.~S.}\ \bibnamefont {Pierce}},
  \bibinfo {author} {\bibfnamefont {A.~K.}\ \bibnamefont {Agrawal}}, \bibinfo
  {author} {\bibfnamefont {P.}~\bibnamefont {Ercius}}, \bibinfo {author}
  {\bibfnamefont {M.}~\bibnamefont {Linck}}, \ and\ \bibinfo {author}
  {\bibfnamefont {B.~J.}\ \bibnamefont {McMorran}},\ }\href {\doibase
  10.1088/1367-2630/16/9/093039} {\bibfield  {journal} {\bibinfo  {journal}
  {New Journal of Physics}\ }\textbf {\bibinfo {volume} {16}},\ \bibinfo
  {pages} {093039} (\bibinfo {year} {2014})},\ \bibinfo {note}
  {00010}\BibitemShut {NoStop}%
\bibitem [{\citenamefont {Grillo}\ \emph {et~al.}(2014)\citenamefont {Grillo},
  \citenamefont {Gazzadi}, \citenamefont {Karimi}, \citenamefont {Mafakheri},
  \citenamefont {Boyd},\ and\ \citenamefont {Frabboni}}]{grillo_highly_2014}%
  \BibitemOpen
  \bibfield  {author} {\bibinfo {author} {\bibfnamefont {V.}~\bibnamefont
  {Grillo}}, \bibinfo {author} {\bibfnamefont {G.~C.}\ \bibnamefont {Gazzadi}},
  \bibinfo {author} {\bibfnamefont {E.}~\bibnamefont {Karimi}}, \bibinfo
  {author} {\bibfnamefont {E.}~\bibnamefont {Mafakheri}}, \bibinfo {author}
  {\bibfnamefont {R.~W.}\ \bibnamefont {Boyd}}, \ and\ \bibinfo {author}
  {\bibfnamefont {S.}~\bibnamefont {Frabboni}},\ }\href {\doibase
  10.1063/1.4863564} {\bibfield  {journal} {\bibinfo  {journal} {Applied
  Physics Letters}\ }\textbf {\bibinfo {volume} {104}},\ \bibinfo {pages}
  {043109} (\bibinfo {year} {2014})}\BibitemShut {NoStop}%
\bibitem [{\citenamefont {Shiloh}\ \emph {et~al.}(2014)\citenamefont {Shiloh},
  \citenamefont {Lereah}, \citenamefont {Lilach},\ and\ \citenamefont
  {Arie}}]{shiloh_sculpturing_2014}%
  \BibitemOpen
  \bibfield  {author} {\bibinfo {author} {\bibfnamefont {R.}~\bibnamefont
  {Shiloh}}, \bibinfo {author} {\bibfnamefont {Y.}~\bibnamefont {Lereah}},
  \bibinfo {author} {\bibfnamefont {Y.}~\bibnamefont {Lilach}}, \ and\ \bibinfo
  {author} {\bibfnamefont {A.}~\bibnamefont {Arie}},\ }\href {\doibase
  10.1016/j.ultramic.2014.04.007} {\bibfield  {journal} {\bibinfo  {journal}
  {Ultramicroscopy}\ }\textbf {\bibinfo {volume} {144}},\ \bibinfo {pages} {26}
  (\bibinfo {year} {2014})},\ \bibinfo {note} {00019}\BibitemShut {NoStop}%
\bibitem [{\citenamefont {Saitoh}\ \emph {et~al.}(2012)\citenamefont {Saitoh},
  \citenamefont {Hasegawa}, \citenamefont {Tanaka},\ and\ \citenamefont
  {Uchida}}]{saitoh_production_2012}%
  \BibitemOpen
  \bibfield  {author} {\bibinfo {author} {\bibfnamefont {K.}~\bibnamefont
  {Saitoh}}, \bibinfo {author} {\bibfnamefont {Y.}~\bibnamefont {Hasegawa}},
  \bibinfo {author} {\bibfnamefont {N.}~\bibnamefont {Tanaka}}, \ and\ \bibinfo
  {author} {\bibfnamefont {M.}~\bibnamefont {Uchida}},\ }\href {\doibase
  10.1093/jmicro/dfs036} {\bibfield  {journal} {\bibinfo  {journal} {Journal of
  Electron Microscopy}\ }\textbf {\bibinfo {volume} {61}},\ \bibinfo {pages}
  {171} (\bibinfo {year} {2012})}\BibitemShut {NoStop}%
\bibitem [{\citenamefont {Schattschneider}\ \emph
  {et~al.}(2012{\natexlab{a}})\citenamefont {Schattschneider}, \citenamefont
  {St{\"o}ger-Pollach},\ and\ \citenamefont
  {Verbeeck}}]{schattschneider_novel_2012}%
  \BibitemOpen
  \bibfield  {author} {\bibinfo {author} {\bibfnamefont {P.}~\bibnamefont
  {Schattschneider}}, \bibinfo {author} {\bibfnamefont {M.}~\bibnamefont
  {St{\"o}ger-Pollach}}, \ and\ \bibinfo {author} {\bibfnamefont
  {J.}~\bibnamefont {Verbeeck}},\ }\href {\doibase
  10.1103/PhysRevLett.109.084801} {\bibfield  {journal} {\bibinfo  {journal}
  {Physical Review Letters}\ }\textbf {\bibinfo {volume} {109}},\ \bibinfo
  {pages} {084801} (\bibinfo {year} {2012}{\natexlab{a}})},\ \bibinfo {note}
  {00035}\BibitemShut {NoStop}%
\bibitem [{\citenamefont {Asenjo-Garcia}\ and\ \citenamefont {Garc{\'i}a~de
  Abajo}(2014)}]{asenjo-garcia_dichroism_2014}%
  \BibitemOpen
  \bibfield  {author} {\bibinfo {author} {\bibfnamefont {A.}~\bibnamefont
  {Asenjo-Garcia}}\ and\ \bibinfo {author} {\bibfnamefont {F.~J.}\ \bibnamefont
  {Garc{\'i}a~de Abajo}},\ }\href {\doibase 10.1103/PhysRevLett.113.066102}
  {\bibfield  {journal} {\bibinfo  {journal} {Physical Review Letters}\
  }\textbf {\bibinfo {volume} {113}},\ \bibinfo {pages} {066102} (\bibinfo
  {year} {2014})},\ \bibinfo {note} {00000}\BibitemShut {NoStop}%
\bibitem [{\citenamefont {Harvey}\ \emph {et~al.}(2015)\citenamefont {Harvey},
  \citenamefont {Pierce}, \citenamefont {Chess},\ and\ \citenamefont
  {McMorran}}]{harvey_demonstration_2015}%
  \BibitemOpen
  \bibfield  {author} {\bibinfo {author} {\bibfnamefont {T.~R.}\ \bibnamefont
  {Harvey}}, \bibinfo {author} {\bibfnamefont {J.~S.}\ \bibnamefont {Pierce}},
  \bibinfo {author} {\bibfnamefont {J.~J.}\ \bibnamefont {Chess}}, \ and\
  \bibinfo {author} {\bibfnamefont {B.~J.}\ \bibnamefont {McMorran}},\ }\href
  {http://arxiv.org/abs/1507.01810} {\bibfield  {journal} {\bibinfo  {journal}
  {arXiv:1507.01810}\ } (\bibinfo {year} {2015})},\ \bibinfo {note} {00001
  arXiv: 1507.01810}\BibitemShut {NoStop}%
\bibitem [{\citenamefont {Idrobo}\ and\ \citenamefont
  {Pennycook}(2011)}]{idrobo_vortex_2011}%
  \BibitemOpen
  \bibfield  {author} {\bibinfo {author} {\bibfnamefont {J.~C.}\ \bibnamefont
  {Idrobo}}\ and\ \bibinfo {author} {\bibfnamefont {S.~J.}\ \bibnamefont
  {Pennycook}},\ }\href {\doibase 10.1093/jmicro/dfr069} {\bibfield  {journal}
  {\bibinfo  {journal} {Journal of Electron Microscopy}\ }\textbf {\bibinfo
  {volume} {60}},\ \bibinfo {pages} {295 } (\bibinfo {year} {2011})},\ \bibinfo
  {note} {00016 Cited by 0004}\BibitemShut {NoStop}%
\bibitem [{\citenamefont {Lloyd}\ \emph {et~al.}(2012)\citenamefont {Lloyd},
  \citenamefont {Babiker},\ and\ \citenamefont {Yuan}}]{lloyd_quantized_2012}%
  \BibitemOpen
  \bibfield  {author} {\bibinfo {author} {\bibfnamefont {S.}~\bibnamefont
  {Lloyd}}, \bibinfo {author} {\bibfnamefont {M.}~\bibnamefont {Babiker}}, \
  and\ \bibinfo {author} {\bibfnamefont {J.}~\bibnamefont {Yuan}},\ }\href
  {\doibase 10.1103/PhysRevLett.108.074802} {\bibfield  {journal} {\bibinfo
  {journal} {Physical Review Letters}\ }\textbf {\bibinfo {volume} {108}},\
  \bibinfo {pages} {074802} (\bibinfo {year} {2012})},\ \bibinfo {note} {00041
  Cited by 0017}\BibitemShut {NoStop}%
\bibitem [{\citenamefont {Schattschneider}\ \emph
  {et~al.}(2012{\natexlab{b}})\citenamefont {Schattschneider}, \citenamefont
  {Schaffer}, \citenamefont {Ennen},\ and\ \citenamefont
  {Verbeeck}}]{schattschneider_mapping_2012}%
  \BibitemOpen
  \bibfield  {author} {\bibinfo {author} {\bibfnamefont {P.}~\bibnamefont
  {Schattschneider}}, \bibinfo {author} {\bibfnamefont {B.}~\bibnamefont
  {Schaffer}}, \bibinfo {author} {\bibfnamefont {I.}~\bibnamefont {Ennen}}, \
  and\ \bibinfo {author} {\bibfnamefont {J.}~\bibnamefont {Verbeeck}},\ }\href
  {\doibase 10.1103/PhysRevB.85.134422} {\bibfield  {journal} {\bibinfo
  {journal} {Physical Review B}\ }\textbf {\bibinfo {volume} {85}},\ \bibinfo
  {pages} {134422} (\bibinfo {year} {2012}{\natexlab{b}})}\BibitemShut
  {NoStop}%
\bibitem [{\citenamefont {Berkhout}\ \emph {et~al.}(2010)\citenamefont
  {Berkhout}, \citenamefont {Lavery}, \citenamefont {Courtial}, \citenamefont
  {Beijersbergen},\ and\ \citenamefont {Padgett}}]{berkhout_efficient_2010}%
  \BibitemOpen
  \bibfield  {author} {\bibinfo {author} {\bibfnamefont {G.~C.~G.}\
  \bibnamefont {Berkhout}}, \bibinfo {author} {\bibfnamefont {M.~P.~J.}\
  \bibnamefont {Lavery}}, \bibinfo {author} {\bibfnamefont {J.}~\bibnamefont
  {Courtial}}, \bibinfo {author} {\bibfnamefont {M.~W.}\ \bibnamefont
  {Beijersbergen}}, \ and\ \bibinfo {author} {\bibfnamefont {M.~J.}\
  \bibnamefont {Padgett}},\ }\href {\doibase 10.1103/PhysRevLett.105.153601}
  {\bibfield  {journal} {\bibinfo  {journal} {Physical Review Letters}\
  }\textbf {\bibinfo {volume} {105}},\ \bibinfo {pages} {153601} (\bibinfo
  {year} {2010})},\ \bibinfo {note} {00231}\BibitemShut {NoStop}%
\bibitem [{\citenamefont {Lavery}\ \emph {et~al.}(2012)\citenamefont {Lavery},
  \citenamefont {Robertson}, \citenamefont {Berkhout}, \citenamefont {Love},
  \citenamefont {Padgett},\ and\ \citenamefont
  {Courtial}}]{lavery_refractive_2012}%
  \BibitemOpen
  \bibfield  {author} {\bibinfo {author} {\bibfnamefont {M.~P.~J.}\
  \bibnamefont {Lavery}}, \bibinfo {author} {\bibfnamefont {D.~J.}\
  \bibnamefont {Robertson}}, \bibinfo {author} {\bibfnamefont {G.~C.~G.}\
  \bibnamefont {Berkhout}}, \bibinfo {author} {\bibfnamefont {G.~D.}\
  \bibnamefont {Love}}, \bibinfo {author} {\bibfnamefont {M.~J.}\ \bibnamefont
  {Padgett}}, \ and\ \bibinfo {author} {\bibfnamefont {J.}~\bibnamefont
  {Courtial}},\ }\href {\doibase 10.1364/OE.20.002110} {\bibfield  {journal}
  {\bibinfo  {journal} {Optics Express}\ }\textbf {\bibinfo {volume} {20}},\
  \bibinfo {pages} {2110} (\bibinfo {year} {2012})},\ \bibinfo {note}
  {00061}\BibitemShut {NoStop}%
\bibitem [{\citenamefont {Malik}\ \emph {et~al.}(2014)\citenamefont {Malik},
  \citenamefont {Mirhosseini}, \citenamefont {Lavery}, \citenamefont {Leach},
  \citenamefont {Padgett},\ and\ \citenamefont {Boyd}}]{malik_direct_2014}%
  \BibitemOpen
  \bibfield  {author} {\bibinfo {author} {\bibfnamefont {M.}~\bibnamefont
  {Malik}}, \bibinfo {author} {\bibfnamefont {M.}~\bibnamefont {Mirhosseini}},
  \bibinfo {author} {\bibfnamefont {M.~P.~J.}\ \bibnamefont {Lavery}}, \bibinfo
  {author} {\bibfnamefont {J.}~\bibnamefont {Leach}}, \bibinfo {author}
  {\bibfnamefont {M.~J.}\ \bibnamefont {Padgett}}, \ and\ \bibinfo {author}
  {\bibfnamefont {R.~W.}\ \bibnamefont {Boyd}},\ }\href {\doibase
  10.1038/ncomms4115} {\bibfield  {journal} {\bibinfo  {journal} {Nature
  Communications}\ }\textbf {\bibinfo {volume} {5}},\ \bibinfo {pages} {3115}
  (\bibinfo {year} {2014})},\ \bibinfo {note} {00055}\BibitemShut {NoStop}%
\bibitem [{\citenamefont {Mirhosseini}\ \emph {et~al.}(2015)\citenamefont
  {Mirhosseini}, \citenamefont {Maga{\~n}a-Loaiza}, \citenamefont
  {O{\textquoteright}Sullivan}, \citenamefont {Rodenburg}, \citenamefont
  {Malik}, \citenamefont {Lavery}, \citenamefont {Padgett}, \citenamefont
  {Gauthier},\ and\ \citenamefont {Boyd}}]{mirhosseini_high-dimensional_2015}%
  \BibitemOpen
  \bibfield  {author} {\bibinfo {author} {\bibfnamefont {M.}~\bibnamefont
  {Mirhosseini}}, \bibinfo {author} {\bibfnamefont {O.~S.}\ \bibnamefont
  {Maga{\~n}a-Loaiza}}, \bibinfo {author} {\bibfnamefont {M.~N.}\ \bibnamefont
  {O{\textquoteright}Sullivan}}, \bibinfo {author} {\bibfnamefont
  {B.}~\bibnamefont {Rodenburg}}, \bibinfo {author} {\bibfnamefont
  {M.}~\bibnamefont {Malik}}, \bibinfo {author} {\bibfnamefont {M.~P.~J.}\
  \bibnamefont {Lavery}}, \bibinfo {author} {\bibfnamefont {M.~J.}\
  \bibnamefont {Padgett}}, \bibinfo {author} {\bibfnamefont {D.~J.}\
  \bibnamefont {Gauthier}}, \ and\ \bibinfo {author} {\bibfnamefont {R.~W.}\
  \bibnamefont {Boyd}},\ }\href {\doibase 10.1088/1367-2630/17/3/033033}
  {\bibfield  {journal} {\bibinfo  {journal} {New Journal of Physics}\ }\textbf
  {\bibinfo {volume} {17}},\ \bibinfo {pages} {033033} (\bibinfo {year}
  {2015})},\ \bibinfo {note} {00066}\BibitemShut {NoStop}%
\bibitem [{\citenamefont {Yan}\ \emph {et~al.}(2014)\citenamefont {Yan},
  \citenamefont {Xie}, \citenamefont {Lavery}, \citenamefont {Huang},
  \citenamefont {Ahmed}, \citenamefont {Bao}, \citenamefont {Ren},
  \citenamefont {Cao}, \citenamefont {Li}, \citenamefont {Zhao}, \citenamefont
  {Molisch}, \citenamefont {Tur}, \citenamefont {Padgett},\ and\ \citenamefont
  {Willner}}]{yan_high-capacity_2014}%
  \BibitemOpen
  \bibfield  {author} {\bibinfo {author} {\bibfnamefont {Y.}~\bibnamefont
  {Yan}}, \bibinfo {author} {\bibfnamefont {G.}~\bibnamefont {Xie}}, \bibinfo
  {author} {\bibfnamefont {M.~P.~J.}\ \bibnamefont {Lavery}}, \bibinfo {author}
  {\bibfnamefont {H.}~\bibnamefont {Huang}}, \bibinfo {author} {\bibfnamefont
  {N.}~\bibnamefont {Ahmed}}, \bibinfo {author} {\bibfnamefont
  {C.}~\bibnamefont {Bao}}, \bibinfo {author} {\bibfnamefont {Y.}~\bibnamefont
  {Ren}}, \bibinfo {author} {\bibfnamefont {Y.}~\bibnamefont {Cao}}, \bibinfo
  {author} {\bibfnamefont {L.}~\bibnamefont {Li}}, \bibinfo {author}
  {\bibfnamefont {Z.}~\bibnamefont {Zhao}}, \bibinfo {author} {\bibfnamefont
  {A.~F.}\ \bibnamefont {Molisch}}, \bibinfo {author} {\bibfnamefont
  {M.}~\bibnamefont {Tur}}, \bibinfo {author} {\bibfnamefont {M.~J.}\
  \bibnamefont {Padgett}}, \ and\ \bibinfo {author} {\bibfnamefont {A.~E.}\
  \bibnamefont {Willner}},\ }\href {\doibase 10.1038/ncomms5876} {\bibfield
  {journal} {\bibinfo  {journal} {Nature Communications}\ }\textbf {\bibinfo
  {volume} {5}},\ \bibinfo {pages} {4876} (\bibinfo {year} {2014})},\ \bibinfo
  {note} {00126}\BibitemShut {NoStop}%
\bibitem [{\citenamefont {Huang}\ \emph {et~al.}(2015)\citenamefont {Huang},
  \citenamefont {Milione}, \citenamefont {Lavery}, \citenamefont {Xie},
  \citenamefont {Ren}, \citenamefont {Cao}, \citenamefont {Ahmed},
  \citenamefont {An~Nguyen}, \citenamefont {Nolan}, \citenamefont {Li},
  \citenamefont {Tur}, \citenamefont {Alfano},\ and\ \citenamefont
  {Willner}}]{huang_mode_2015}%
  \BibitemOpen
  \bibfield  {author} {\bibinfo {author} {\bibfnamefont {H.}~\bibnamefont
  {Huang}}, \bibinfo {author} {\bibfnamefont {G.}~\bibnamefont {Milione}},
  \bibinfo {author} {\bibfnamefont {M.~P.~J.}\ \bibnamefont {Lavery}}, \bibinfo
  {author} {\bibfnamefont {G.}~\bibnamefont {Xie}}, \bibinfo {author}
  {\bibfnamefont {Y.}~\bibnamefont {Ren}}, \bibinfo {author} {\bibfnamefont
  {Y.}~\bibnamefont {Cao}}, \bibinfo {author} {\bibfnamefont {N.}~\bibnamefont
  {Ahmed}}, \bibinfo {author} {\bibfnamefont {T.}~\bibnamefont {An~Nguyen}},
  \bibinfo {author} {\bibfnamefont {D.~A.}\ \bibnamefont {Nolan}}, \bibinfo
  {author} {\bibfnamefont {M.-J.}\ \bibnamefont {Li}}, \bibinfo {author}
  {\bibfnamefont {M.}~\bibnamefont {Tur}}, \bibinfo {author} {\bibfnamefont
  {R.~R.}\ \bibnamefont {Alfano}}, \ and\ \bibinfo {author} {\bibfnamefont
  {A.~E.}\ \bibnamefont {Willner}},\ }\href {\doibase 10.1038/srep14931}
  {\bibfield  {journal} {\bibinfo  {journal} {Scientific Reports}\ }\textbf
  {\bibinfo {volume} {5}} (\bibinfo {year} {2015}),\ 10.1038/srep14931},\
  \bibinfo {note} {00025}\BibitemShut {NoStop}%
\bibitem [{\citenamefont {Hossack}\ \emph {et~al.}(1987)\citenamefont
  {Hossack}, \citenamefont {Darling},\ and\ \citenamefont
  {Dahdouh}}]{hossack_coordinate_1987}%
  \BibitemOpen
  \bibfield  {author} {\bibinfo {author} {\bibfnamefont {W.~J.}\ \bibnamefont
  {Hossack}}, \bibinfo {author} {\bibfnamefont {A.~M.}\ \bibnamefont
  {Darling}}, \ and\ \bibinfo {author} {\bibfnamefont {A.}~\bibnamefont
  {Dahdouh}},\ }\href {\doibase 10.1080/09500348714551121} {\bibfield
  {journal} {\bibinfo  {journal} {Journal of Modern Optics}\ }\textbf {\bibinfo
  {volume} {34}},\ \bibinfo {pages} {1235} (\bibinfo {year} {1987})},\ \bibinfo
  {note} {00031}\BibitemShut {NoStop}%
\bibitem [{\citenamefont {Matteucci}\ \emph {et~al.}(1992)\citenamefont
  {Matteucci}, \citenamefont {Missiroli}, \citenamefont {Muccini},\ and\
  \citenamefont {Pozzi}}]{matteucci_electron_1992}%
  \BibitemOpen
  \bibfield  {author} {\bibinfo {author} {\bibfnamefont {G.}~\bibnamefont
  {Matteucci}}, \bibinfo {author} {\bibfnamefont {G.~F.}\ \bibnamefont
  {Missiroli}}, \bibinfo {author} {\bibfnamefont {M.}~\bibnamefont {Muccini}},
  \ and\ \bibinfo {author} {\bibfnamefont {G.}~\bibnamefont {Pozzi}},\ }\href
  {\doibase 10.1016/0304-3991(92)90039-M} {\bibfield  {journal} {\bibinfo
  {journal} {Ultramicroscopy}\ }\textbf {\bibinfo {volume} {45}},\ \bibinfo
  {pages} {77} (\bibinfo {year} {1992})},\ \bibinfo {note} {00043}\BibitemShut
  {NoStop}%
\bibitem [{\citenamefont {Cumings}\ \emph {et~al.}(2002)\citenamefont
  {Cumings}, \citenamefont {Zettl}, \citenamefont {McCartney},\ and\
  \citenamefont {Spence}}]{cumings_electron_2002}%
  \BibitemOpen
  \bibfield  {author} {\bibinfo {author} {\bibfnamefont {J.}~\bibnamefont
  {Cumings}}, \bibinfo {author} {\bibfnamefont {A.}~\bibnamefont {Zettl}},
  \bibinfo {author} {\bibfnamefont {M.~R.}\ \bibnamefont {McCartney}}, \ and\
  \bibinfo {author} {\bibfnamefont {J.~C.~H.}\ \bibnamefont {Spence}},\ }\href
  {\doibase 10.1103/PhysRevLett.88.056804} {\bibfield  {journal} {\bibinfo
  {journal} {Physical Review Letters}\ }\textbf {\bibinfo {volume} {88}},\
  \bibinfo {pages} {056804} (\bibinfo {year} {2002})},\ \bibinfo {note}
  {00121}\BibitemShut {NoStop}%
\bibitem [{\citenamefont {Beleggia}\ \emph {et~al.}(2014)\citenamefont
  {Beleggia}, \citenamefont {Kasama}, \citenamefont {Larson}, \citenamefont
  {Kelly}, \citenamefont {Dunin-Borkowski},\ and\ \citenamefont
  {Pozzi}}]{beleggia_towards_2014}%
  \BibitemOpen
  \bibfield  {author} {\bibinfo {author} {\bibfnamefont {M.}~\bibnamefont
  {Beleggia}}, \bibinfo {author} {\bibfnamefont {T.}~\bibnamefont {Kasama}},
  \bibinfo {author} {\bibfnamefont {D.~J.}\ \bibnamefont {Larson}}, \bibinfo
  {author} {\bibfnamefont {T.~F.}\ \bibnamefont {Kelly}}, \bibinfo {author}
  {\bibfnamefont {R.~E.}\ \bibnamefont {Dunin-Borkowski}}, \ and\ \bibinfo
  {author} {\bibfnamefont {G.}~\bibnamefont {Pozzi}},\ }\href {\doibase
  10.1063/1.4887448} {\bibfield  {journal} {\bibinfo  {journal} {Journal of
  Applied Physics}\ }\textbf {\bibinfo {volume} {116}},\ \bibinfo {pages}
  {024305} (\bibinfo {year} {2014})},\ \bibinfo {note} {00003}\BibitemShut
  {NoStop}%
\bibitem [{\citenamefont {McCartney}\ and\ \citenamefont
  {Smith}(2007)}]{mccartney_electron_2007}%
  \BibitemOpen
  \bibfield  {author} {\bibinfo {author} {\bibfnamefont {M.~R.}\ \bibnamefont
  {McCartney}}\ and\ \bibinfo {author} {\bibfnamefont {D.~J.}\ \bibnamefont
  {Smith}},\ }\href {\doibase 10.1146/annurev.matsci.37.052506.084219}
  {\bibfield  {journal} {\bibinfo  {journal} {Annual Review of Materials
  Research}\ }\textbf {\bibinfo {volume} {37}},\ \bibinfo {pages} {729}
  (\bibinfo {year} {2007})},\ \bibinfo {note} {00116}\BibitemShut {NoStop}%
\bibitem [{\citenamefont {Feynman}(1964)}]{feynman_chapter_1964}%
  \BibitemOpen
  \bibfield  {author} {\bibinfo {author} {\bibfnamefont {R.~P.}\ \bibnamefont
  {Feynman}},\ }in\ \href@noop {} {{\selectlanguage {english}\emph {\bibinfo
  {booktitle} {The {Feynman} {Lectures} on {Physics}}}}},\ Vol.~\bibinfo
  {volume} {2}\ (\bibinfo  {publisher} {Addison-Wesley},\ \bibinfo {address}
  {Reading, MA},\ \bibinfo {year} {1964})\BibitemShut {NoStop}%
\end{thebibliography}%

\appendix
\begin{widetext}

\section{Crosstalk}
An important figure of merit for a measurement device is the crosstalk: the rate of erroneous counts that occur when adjacent measurement outcomes are counted as the outcome of interest. Figure \ref{fig:crosstalk} shows the the crosstalk of an ideal electron orbital angular momentum sorter, simulated with phases shown in \ref{eq:vpend} and \ref{eq:corrector_phase} and parameters shown in Table \ref{table:params}.

\begin{figure}[h]
  \subfloat[]{\label{subfig:crosstalk}
  \includegraphics[height=2.5in]{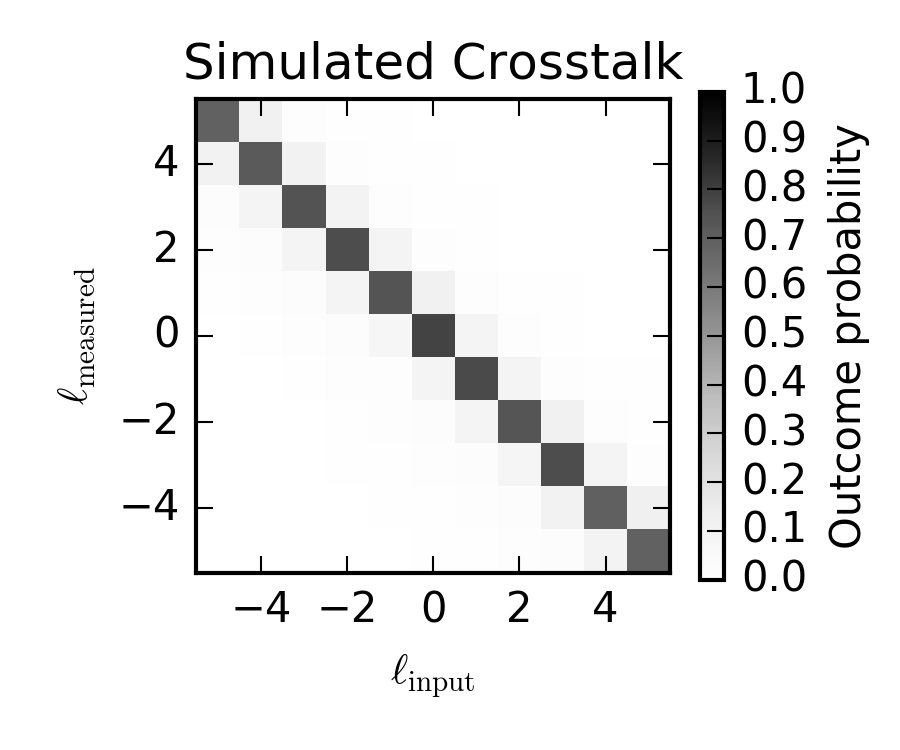}}
  \caption{Crosstalk of the electron orbital angular momentum measurement with parameters shown in Table \ref{table:params}. A perfect sorter would have outcome probabilities of exactly 1 for every $\ell_{\mathrm{measured}} = \ell_{\mathrm{input}}$ and 0 elsewhere. \label{fig:crosstalk}}
\end{figure}

\section{Calculation of phase past charged needle}
Here we consider electrons propagating in the $z$ direction past an infinitesimally thin needle of constant charge density $\sigma= Q/L$, where $L$ is the length of the needle. We consider that the needle lies on the $-y'$-axis with one tip at the origin and the other located at $y = -L$. The charged needle is oriented perpendicularly to a conducting plate that lies parallel to the $x-z$ plane at $y = h$. The electrostatic potential of this arrangement is

\begin{equation}
  \begin{aligned}
  V(\mathbf{r}) =& \frac{Q}{4\pi\epsilon_0 L} \ln\Bigg[\left(\frac{y - h - L + \sqrt{x^2 + (y - h - L)^2+z^2}}{y - h + \sqrt{x^2+(y - h)^2+z^2}}\right) \\
  \times & \left(\frac{y - L + \sqrt{x^2+(y - L)^2+z^2}}{y + \sqrt{x^2+y^2+z^2}}\right)\Bigg]
  \end{aligned}
    \label{eq:V}
\end{equation}

Following \cite{matteucci_electron_1992}, we use Eq. \ref{eq:vp1a} to calculate the phase an electron plan wave acquires as it propagates through this potential:

\begin{equation}
  \begin{aligned}
	\varphi(\mathbf{r})  = & \frac{Q C_E}{2 \pi \epsilon_0 L}\Bigg[ x \sin ^{-1}\left(\frac{y+2 h}{\sqrt{x^2+(y+2 h)^2}}\right) - x \sin ^{-1}\left(\frac{y - L+2 h}{\sqrt{x^2 + (y - L+2 h)^2}}\right) \\
	& + x \sin ^{-1} \left(\frac{y-L}{\sqrt{(y-L)^2+x^2}}\right) - x \sin^{-1}\left(\frac{y}{\sqrt{x^2+y^2}}\right)\\
	&-y \ln \left( \frac{\sqrt{x^2+y^2}}{\sqrt{x^2+(y-L)^2}}\right)  + y \ln \left( \frac{\sqrt{x^2+(y+2 h)^2}}{\sqrt{x^2+(y - L + 2 h)^2}} \right)\\
	&- L \ln \left(\frac{\sqrt{x^2+(y-L)^2}}{\sqrt{x^2+(y - L + 2 h)^2}}\right) + 2 h \ln \left(\frac{\sqrt{x^2+(y + 2 h)^2}}{\sqrt{x^2+(y - L + 2 h)^2}}\right) \Bigg].
  \end{aligned}
    \label{eq:vp1d}
\end{equation}

We consider a situation in which the incident electron beam is confined only to the region $(x,y)$ immediately surrounding the tip of the needle nearest to the plate. Taking the limit of Eq. \ref{eq:vp1d} as the distance $h$ between the needle and the plate goes to infinity, we see that the first two terms cancel, the sixth and seventh terms go to zero, and the last term goes to a constant (albeit infinite) phase shift:

\begin{equation}
  \begin{aligned}
	\varphi(\mathbf{r})  = & \frac{Q C_E}{2 \pi \epsilon_0 L}\Bigg[ x \sin ^{-1} \left(\frac{y-L}{\sqrt{x^2+(y-L)^2}}\right) - x \sin^{-1}\left(\frac{y}{\sqrt{x^2+y^2}}\right)\\
	&-y \ln \left( \frac{\sqrt{x^2+y^2}}{\sqrt{x^2+(y-L)^2}}\right) \Bigg] + \varphi_0,
  \end{aligned}
    \label{eq:vp1e}
\end{equation}
\noindent where $\varphi_0$ is a constant uniform ``background" phase that is experimentally unobservable.

We now assume that the length of the needle is large compared to the region of interest, such that $L \gg \sqrt{x^2 + y^2}$, although finite in extent. With this approximation, the phase distribution induced onto an electron wave passing close to the nearest tip of the needle is:

\begin{equation}
  \begin{aligned}
	\varphi_{\textrm{tip}}(x,y) &= \frac{Q C_E}{2 \pi \epsilon_0 L}\left[ x \sin ^{-1}\left(\frac{x}{\sqrt{x^2+y^2}}\right) + y \ln \left(\frac{\sqrt{x^2+y^2}}{L}\right)\right] + \varphi_0 \\
	&= \frac{Q}{2 \epsilon_0 L \lambda E}\left[ x \arctan{\left(\frac{x}{y}\right)} + y \ln \left(\frac{\sqrt{x^2+y^2}}{L}\right)\right] + \varphi_0.
  \end{aligned}
    \label{eq:vptip}
\end{equation}

Eq. \ref{eq:vptip} is exactly the desired phase of the unwrapper element (Eq. \ref{eq:phi1}) minus a linear phase that can easily be applied electrostatically by adjusting the bias of two charged plates.

Note that Eq. (4) in \cite{matteucci_electron_1992} does not make the approximations above, and describes the total phase imprinted on an electron by the full potential of both ends of the needle, as well as those of its image charge in the plate. However, note that Matteucci \emph{et al.}'s Equation 4 can be expressed directly in terms of Eq. \ref{eq:vptip}:

\begin{equation}
  \begin{aligned}
  \varphi(\mathbf{r'}) & = \varphi_{\textrm{tip}}(x',y' + c + h) \\
  & +\varphi_{\textrm{tip}}(x',y' - c - h) \\
  & +\varphi_{\textrm{tip}}(x',-y' - c + h) \\
  & +\varphi_{\textrm{tip}}(x',-y' + c - h).
  \end{aligned}
    \label{eq:vp1b}
\end{equation}

\noindent
Where here we adopting their notation by substituting $L = 2 c$, and use shifted, primed coordinates (x' = x, y' = y + c - h, z' = z). This form reveals that the total phase calculated by Matteucci \emph{et al.} can be interpreted as a sum of four individual phases induced by each of the two ends of the needle as well as the ends of the ``image" of the needle within the plate electrode. 

\section{Phase of corrector element \label{sect:corrector_depth}}

As the corrector phase solves Laplace's equation, i.e. $\nabla^2 \varphi_c(u,v) = 0$, it is straightforward to generate this phase with an electrostatic potential $V(u,v)$, following Eq. \ref{eq:vp1a}. We can approximate the two-dimensional solution to Laplace's equation $V(u,v)$ with a nearly-$z$-independent three-dimensional solution. The simplest boundary conditions are constant over a range in $z$ that we'll call the depth, $D$. In particular, we can specify the $V(u,z)$ we want with boundaries at $u = 0$ and $u=u_1$. In other words,
\begin{equation}
  V(u_i,v,z) = \begin{cases} 
    V(u_i,v) & \qquad |z| \leq \frac{D}{2} \\
    \mathrm{free} & \qquad \mathrm{elsewhere}
  \end{cases}
\end{equation}

We investigated these boundaries with a numerical solution to Laplace's equation. In the range $|z| < \frac{D}{2}$, we set the Dirichlet boundary conditions
\begin{align}
  V(u=0,v) &= V_{c_0} \cos\left(-\frac{2 \pi v}{d}\right) \\
  V(u=u_0,v) &= V_{c_1} \cos\left(-\frac{2 \pi v}{d}\right),
\end{align}
where $V_{c_0}$ and $V_{c_1}$ are the peak potentials at $u=u_0 = 0$ and $u=u_1$, respectively, and $d$ is the period in $v$. We see that, to satisfy Laplace's equation, we must have $V_{c_1} = V_{c_0} \exp{\left(-\frac{2 \pi u_1}{d}\right)}$. We used peridoic bondary conditions in $v$, and the von Neumann boundary condition $\boldsymbol{\nabla} V \cdot \hat{\mathbf{n}} = 0$ for all other boundaries.
\begin{figure}[h]
  \subfloat[]{\label{subfig:corrector:x-z_slice}
  \includegraphics[width=3in]{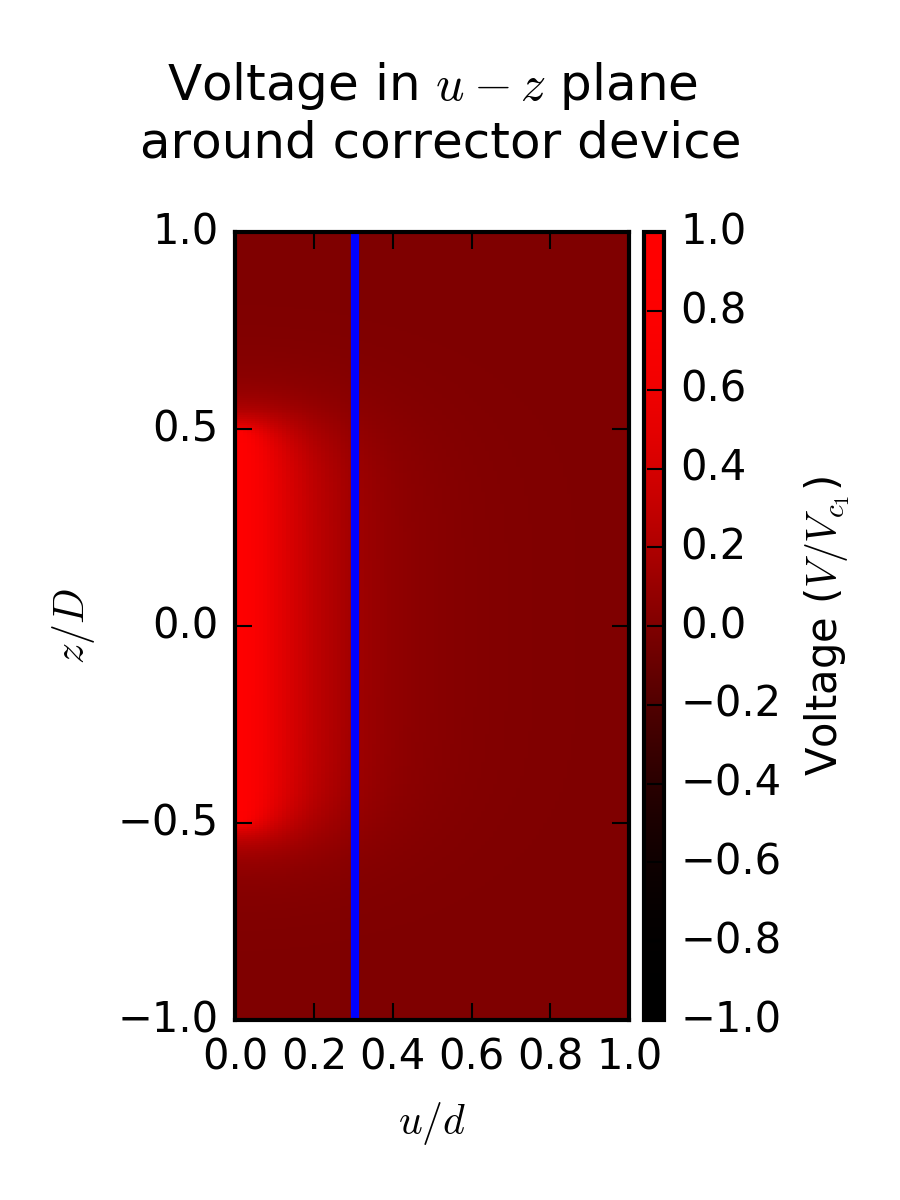} }
  \subfloat[]{\label{subfig:corrector:z_profile}
  \includegraphics[width=4in]{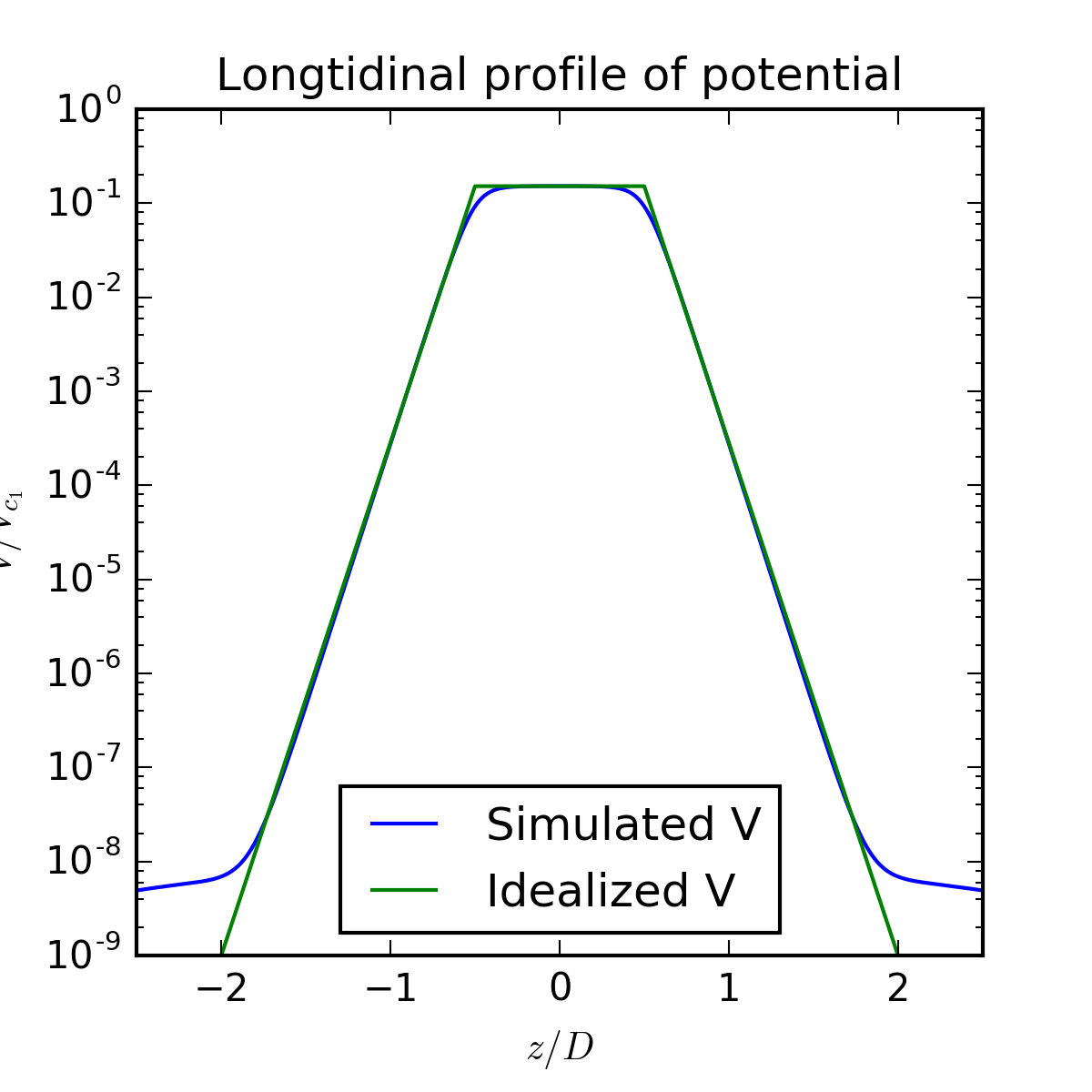} }
  \caption{(a) Cross-section of simulated potential in the $u$-$z$ plane at $v=0$ showing rapid decay of potential outside the device. \\
  (b) Line plot of a simulated potential at $u = 0.3 d$ in the $u$-$z$ plane showing exponential decay of the potential outside the device. (blue) Simulated potential $V(u = 0.3 d,v=0,z)$, also shown as blue line in (a); (green) Model of the potential that is constant inside the device and exponentially decays as $V \propto \exp\left(\pm\frac{2\pi (z\pm D/2)}{d}\right)$ outside the device. \\
This simulation used a period $d = 1.0$, a depth $D = 2.0$ (resulting in boundaries at $z=\pm 1.0$), arbitary $V_{c_0}$, and boundaries at $u = 0$, $u=1.0$, $v = 0$, $v = 1.0$, $z = -50.0$ and $z = 50.0$ with a voxel size of $0.01$ by $2^{-5}$ by $0.01$. \\
}
\end{figure}
We found that, as long as the depth $D$ was much larger than the period $d$, i.e. the potential is constant in $z$ over a much longer length scale than it varies in $u$ and $v$, the fringing fields were insignificant. Specifically, we found that the potential decayed exponentially with a decay length $\frac{d}{2\pi}$ outside the device. The contribution of this tail to the phase scales with $d$, while the contribution from inside the device scales with $D$. The precision of the phase can therefore be arbitrarily increased by increasing $D$ while holding $d$ constant, up to the limit of the thin grating condition $\lambda D \ll d^2$. As $\lambda = \unit[1.97]{\textrm{pm}}$ for $\unit[300]{\textrm{keV}}$ electrons, if $d = \unit[1]{\textrm{mm}}$, the device would still act as a thin grating up to $D \sim \unit[10^6]{\textrm{m}}$. 

\begin{figure}
  \subfloat[]{\label{subfig:corrector:x-y_slice_sinusoidal}
  \includegraphics[width=3.5in]{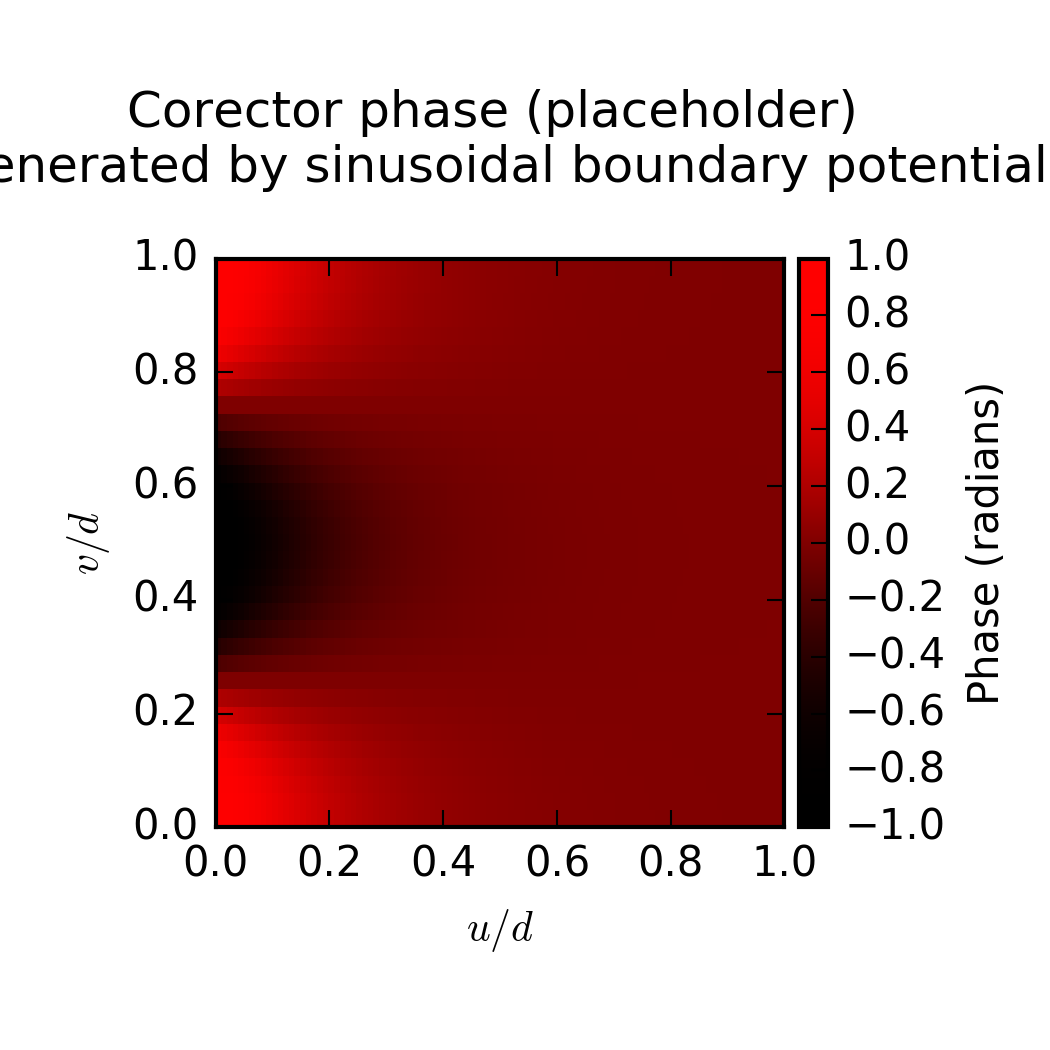} }
  \subfloat[]{\label{subfig:corrector:x-y_slice_flat}
  \includegraphics[width=3.5in]{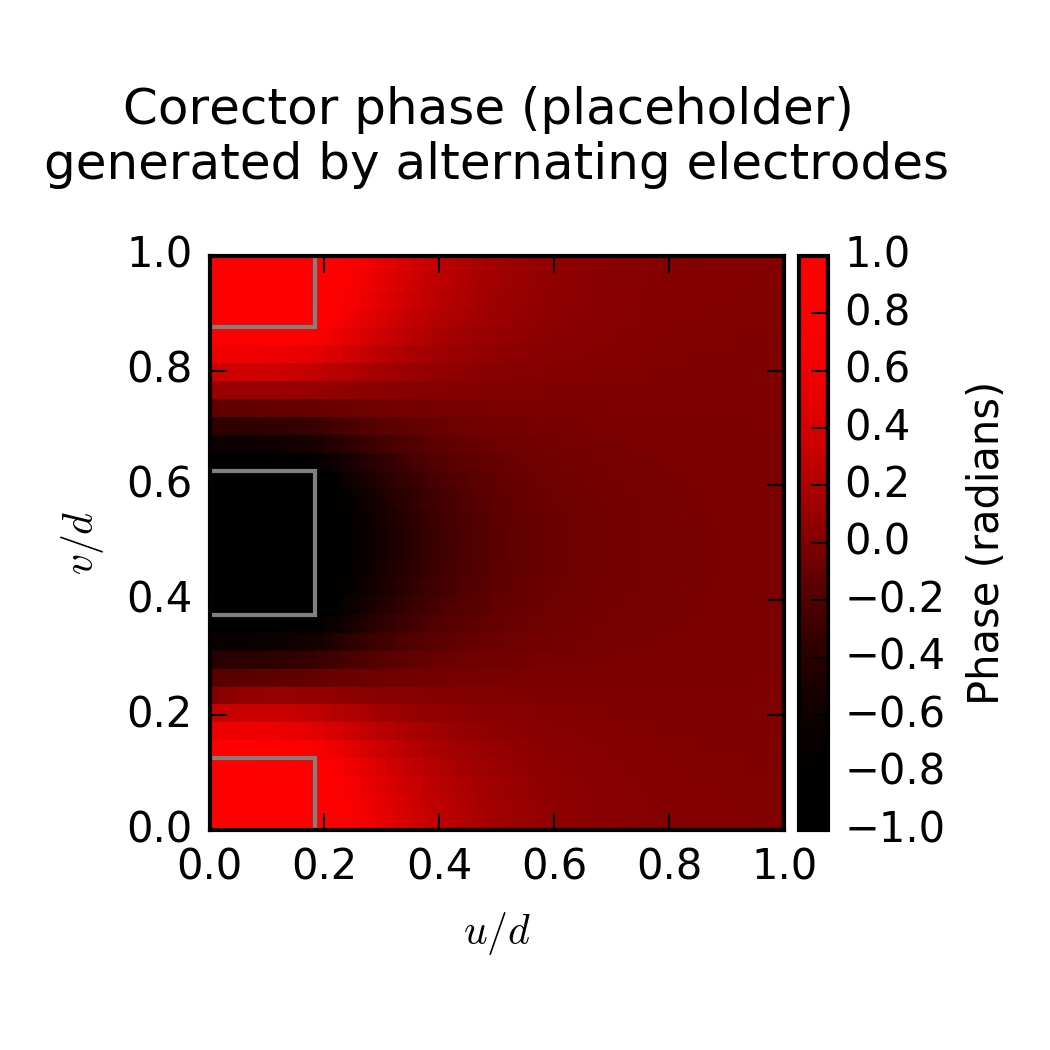}}
  \caption{Simulated phases (potentials). (a) with sinusoidal BCs, (b) with flat electrodes. (grey) outline of electrode.}
\end{figure}

\end{widetext}
\end{document}